\documentclass[journal]{IEEEtran}
\pdfoutput=1
\usepackage{setspace}
\usepackage{indentfirst}
\usepackage[numbers,sort&compress]{natbib}
\usepackage{color}
\usepackage[pdftex]{graphicx}
\usepackage{amsmath}
\usepackage{amsthm}
\usepackage{amssymb}
\usepackage[ruled]{algorithm2e}
\usepackage{array}
\usepackage[caption=true,font=footnotesize]{subfig}
\usepackage{fixltx2e}
\usepackage{stfloats}
\usepackage{url}
\usepackage{comment}
\usepackage{color}
\usepackage[normalem]{ulem}

\newtheorem{thm}{Theorem}
\newtheorem{lem}[thm]{Lemma}

\newtheorem{cor}{Corollary}
\theoremstyle{definition}
\newtheorem{defn}{Definition}

\newcommand{\tabincell}[2]{\begin{tabular}{@{}#1@{}}#2\end{tabular}}

\begin{document}

\title{Hierarchical Multi-resource Fair Queueing for Packet Processing}
\author{Chaoqun~You,~\IEEEmembership{Member,~IEEE,} Yangming~Zhao, Gang Feng,~\IEEEmembership{Senior Member,~IEEE,}\\ Tony Q. S. Quek,~\IEEEmembership{Fellow,~IEEE,} and Lemin~Li
\thanks{This work was supported in part by the Singapore Ministry of Education (MOE) Academic Research Fund (AcRF) Tier 2 under Grant T2EP20120-0006, in part by the Anhui Provincial Natural Science Foundation No. 2208085MF167, in part by the Natural Science Foundation of Jiangsu Province No. BK20221261, and in part by the Hefei Municipal Natural Science Foundation No. 2022004. \textit{(Corresponding author: Yangming Zhao)} }
\thanks{This paper is a substantially extended version of the paper ``Hierarchical Multi-resource Fair Queueing for Network Function Virtualization" at IEEE INFOCOM 2019~\cite{chaoqun2019hierarchical}.}
\thanks{C. You and T. Q. S. Quek are with the Wireless Networks and Decision Systems group, Singapore University of Technology and Design, Singapore 487372 (e-mail:chaoqun\_you, tonyquek@sutd.edu.sg).}
\thanks{Y. Zhao is with the with the School of Computer Science and Technology, University of Science and Technology of China, Hefei 611731, China (e-mail: zhaoym.ustc@gmail.com).}
\thanks{G. Feng and L. Li are with the
University of Electronic Science and Technology of China, Chengdu
611731, China (e-mail: fenggang, lml@uestc.edu.cn).}}

\maketitle
\begin{abstract}
Various middleboxes are ubiquitously deployed in networks to perform packet processing functions, such as firewalling, proxy, scheduling, etc., for the flows passing through them. With the explosion of network traffic and the demand for multiple types of network resources, it has never been more challenging on a middlebox to provide Quality-of-Service (QoS) guarantees to grouped flows. Unfortunately, all currently existing fair queueing algorithms fail in supporting hierarchical scheduling, which is necessary to provide QoS guarantee to the grouped flows of multiple service classes. In this paper, we present two new multi-resource fair queueing algorithms to support hierarchical scheduling, collapsed Hierarchical Dominant Resource Fair Queueing (collapsed H-DRFQ) and dove-tailing H-DRFQ. Particularly, collapsed H-DRFQ transforms the hierarchy of grouped flows into a flat structure for flat scheduling while dove-tailing H-DRFQ iteratively performs flat scheduling to sibling nodes on the original hierarchy. Through rigorous theoretical analysis, we find that both algorithms can provide hierarchical share guarantees to individual flows, while the upper bound of packet delay in dove-tailing H-DRFQ is smaller than that of collapsed H-DRFQ. We implement the proposed algorithms on Click modular router and the experimental results verify our analytical results. 
\end{abstract}


\maketitle
\IEEEpeerreviewmaketitle

\section{Introduction}\label{sec:1}

\IEEEPARstart{M}{iddleboxes}, also known as network functions, ranging from firewalling, proxies to Wide Area Network (WAN) optimizers, are ubiquitously deployed in networks to perform packet processing functions on the traffic flows passing through them~\cite{ren2020untold},~\cite{newman2020control}.
With the development of network virtualization technology, there have been increasing proposals on software-defined middlebox,  which suggest consolidating more functions onto the same device, thereby further expanding the scope of middlebox applications.
For example, network function virtualization (NFV) moves packet processing from dedicated hardware middleboxes to virtualized network functions (VNFs) running on commodity (\emph{e.g.}, x86 based systems) servers~\cite{liu2018microboxes}.
By decoupling the VNF instances from the underlying hardware, NFV technology enables network operators to readily deploy new services to existing middleboxes~\cite{zheng2020sfc}, \cite{zheng2020orchestrating}.

As the volume of traffic surges~\cite{honda2011still}, it is increasingly important to provide Quality-of-Service (QoS) guarantees across traffic flows on middleboxes.
Fair queueing is a fundamental mechanism to achieve this objective such that the performance \emph{isolation} among flows can be guaranteed.
With fair queueing, a scheduler determines the order to forward packets from multiple flows on a shared resource, and allocates a prescribed fair share to individual flows~\cite{parekh1993generalized}.
%
%
Achieving fair queueing in middleboxes, however, is particularly challenging, due to the traffic aggregation and diversified resource demands.

Particularly, traffic flows are grouped to support multiple service classes that may include real-time service, best-time service and others~\cite{bennett1997hierarchical}.
A group is formed by aggregating a set of flows according to administrative affiliation, protocol, traffic type, \emph{etc}~\cite{bennett1997hierarchical},~\cite{stoica2000hierarchical}.
A multi-level tree (also referred as hierarchy) can be used to describe the organizational structure of traffic flows, with its leaves representing individual flows and its internal nodes representing groups.
In such a multi-group scenario, it is desirable for the scheduler to guarantee performance isolation, not only among the flows but also among groups.

Traffic flows are also characterized by the degree of demand diversity across \emph{multiple} types of resources. Different network functions consume vastly different amount and types of resources. For example, basic forwarding functionality uses more link bandwidth than other resources. IPSec encryption, on the other hand, is CPU-consuming. Indeed, payload analyses on production traces from Intel, Google and Facebook~\cite{ghodsi2012multi},~\cite{ghodsi2011dominant},~\cite{reiss2012heterogeneity} confirm that flows may consume vastly different amount of resources of CPU, memory, I/O bandwidth and network bandwidth.

Extensive existing works have concentrated on multi-resource fair queuing recently.
Specifically, Ghodsi \emph{et al}.~\cite{ghodsi2012multi} proposed the first work on multi-resource fair queueing, named Dominant Resource Fairness Queueing (DRFQ).
It ensures that the flows receive roughly the same packet processing time on their respective \emph{dominant resources}, \emph{i.e.}, the resources they respectively require the most processing time on.
From DRFQ, numerous investigations have been proposed to improve the multi-resource fair queueing policies~\cite{wang2013multi2},~\cite{wang2013multi},~\cite{wang2014low},~\cite{li2015low},~\cite{chen2017cluster}.
Unfortunately, thus far there are no existing fair queueing algorithms which can support hierarchies in group scheduling.
In fact, with hierarchies, even the packet order of internal nodes is \emph{unclear}: with a hierarchical scheduler, the queue at an internal node is a logical one, and its order of packets is determined by the scheduling policy applied to its children.
Moreover, naive extensions of DRFQ to support hierarchies violate the hierarchical share guarantees, which is a key property to ensure each node in the hierarchy gets at least its prescribed fair share regardless of other nodes' behaviors (which will be proven in Section~\ref{sec:3}).

In this paper, we focus on multi-resource fair queueing policy to support hierarchical scheduling. To the best of our knowledge, this is the first rigorous study on fair queueing that supports hierarchies in multi-resource middleboxes.
We propose two new hierarchical multi-resource fair queueing algorithms, namely collapsed Hierarchical Dominant Resource Fair Queueing (collapsed H-DRFQ) and dove-tailing H-DRFQ, to determine the packet orders of grouped flows on middlebox.
The main idea of collapsed H-DRFQ is to convert the hierarchy into a flat tree and then use the flattened tree as a DRFQ scheduler, while dove-tailing H-DRFQ follows the idea to execute DRFQ within each set of sibling nodes (the nodes that share the same parent).

Besides, the proposed algorithms satisfy two desired properties. First, H-DRFQ provides \emph{hierarchical share guarantee}: a node in a hierarchy is guaranteed to get its desirable fair share of resources, regardless of the behaviors of others. Second, H-DRFQ is \emph{group-strategyproof}: no group of flows can increase their allocation by inflating their resource demands.

To evaluate our proposed H-DRFQ algorithms, we analyse the packet scheduling delay through both analytical modeling and experiments based on Click modular router.
Theoretical analysis shows that dove-tailing H-DRFQ provides a smaller delay bound than collapsed H-DRFQ. However, according to the experimental results, both algorithms have their pros and cons. Collapsed H-DRFQ is more appropriate for the case of simple structures while dove-tailing H-DRFQ has advantages for the case of complicated hierarchies.

To summarize, our contributions in this work are five-fold:
\begin{itemize}
  \item We identify the problem of \emph{hierarchical} multi-resource fair queueing.
  \item We propose two hierarchical multi-resource fair queueing algorithms, collapsed H-DRFQ and dove-tailing H-DRFQ. (Section~\ref{sec:4})
  \item We theoretically prove the existence of the upper bound of the packet scheduling delay for the two H-DRFQ algorithms respectively. (Section~\ref{sec:5})
  \item We prove both H-DRFQ algorithms satisfy the properties of providing hierarchical share guarantees and being group strategy-proofness. (Section~\ref{sec:6})
  \item We implement the algorithms on a testbed based on Click modular router and conduct extensive simulations to evaluate the performance of H-DRFQ. (Section~\ref{sec:7} and Section~\ref{sec:8})
\end{itemize}

\section{Preliminaries} \label{sec:2}

In this section, we introduce the notations and some essential properties of hierarchical scheduling that are required for discussing hierarchial multi-resources fair queueing problem. Then we review the packet scheduling scheme, DRFQ, which is the base for our proposals.

\subsection{Notations} \label{sec:2.1}
\begin{figure}[t!]
\centering
  \includegraphics[width=2.0 in]{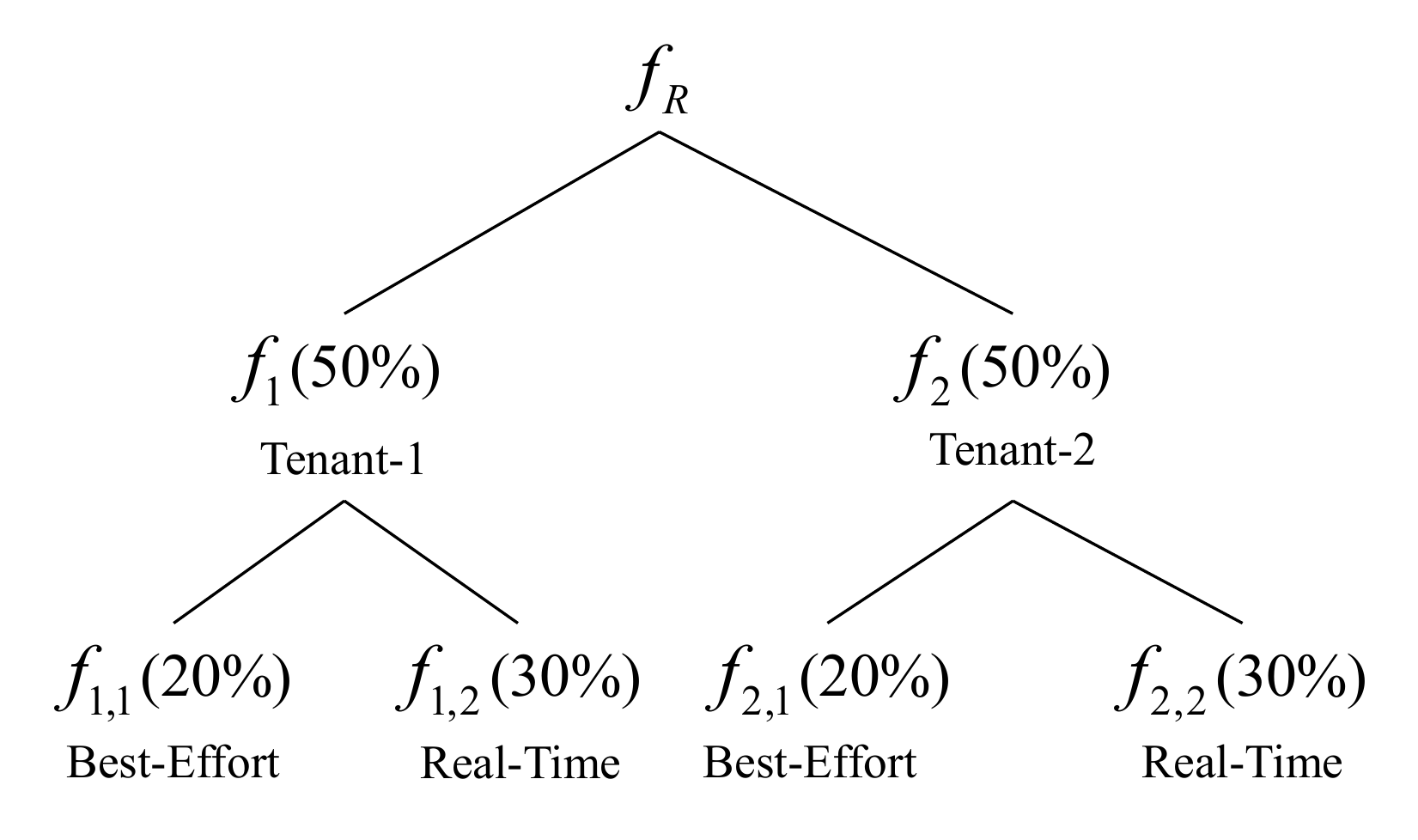}
  \caption{Example hierarchy, with weights in parenthesis}
  \label{fig_9}
\end{figure}

Hierarchical scheduling is usually performed in the buffers at output ports of middleboxes, in which an output buffer is the place where packets ``enter" a link.
A hierarchical scheduler is specified with a weighted tree (see Fig.~\ref{fig_9}).
The root node represents a VNF scheduler and a leaf node represents a flow to be served by the scheduler.
Each internal node represents a group that may contain multiple sub-groups or flows.
Fig.~\ref{fig_9} shows an example where there are two internal nodes $f_1$ and $f_2$ sharing the VNF resources, each of which represents a tenant that contains two type of flows with different service classes, best-effort and real-time.

A node on the tree is denoted by $f_i$, where $i$ is a list of numbers that describe how the node can be found starting from the top of the tree and going down, left to right. For example, $f_{2,1}$ is found by starting at the root, picking its second (from left) child, and picking that child's first child. The root node is $f_R$ (or $R$). Each flow $i$ is associated with a weight $\phi_i$. Without loss of generality, we assume that the sum of weights of all leaf nodes in the hierarchy is normalized to 1. The parent of flow $i$ is denoted by $P(i)$, the set of children of flow $i$ is denoted by $C(i)$, and the set of sibling nodes of flow $i$ is denoted by $Sib(i)$. Then we have $\sum_{j\in C(i)} \phi_j = \phi_i$. For each flow $i$ with $H$ ancestors in a hierarchy, we use $P^h(i)$ to represent its $h$th predecessor for $h = 0,1,\dots,H$, where $P^1(i) = P(i)$ and $P^H(i)=R$.

Let $m$ be the number of resource types. Denote the $k$th packet of flow $i$ as $p_i^k$. The \emph{dominant resource} of a packet is defined as the one that requires the maximum packet processing time. Specifically, let $s_{i,r}^k$ be the packet processing time of $p_i^k$ on resource $r$, the dominant resource of $p_i^k$ is $r_{p_i^k} = \arg\max_r s_{i,r}^k$. For example, assume there are two types of resources, CPU and bandwidth, in a middlebox. If the processing time requirement of a packet passing through this middlebox is $\langle 2,1 \rangle$, which means this packet requires 2 units of packet processing time on CPU and 1 unit of packet processing time on bandwidth, then the dominant resource of this packet is CPU. We refer the processing time requirements $\{s_{i,r}^k\}$ ($r=1,\dots,m$) of a packet $p_i^k$ as its \emph{packet profile}. And we denote $\mu_{i}^k$ as the processing time of $p_{i}^k$ on its dominant resource (\emph{i.e.}, $\mu_i^k = \max_r s_{i,r}^k$). Moreover, we denote $W_i(t_1, t_2)$ as the total processing time consumed by flow $i$ on its dominant resource during time interval $[t_1,t_2]$.

\subsection{Properties of Hierarchical Scheduling} \label{sec:2.2}

Previous work on DRFQ identifies several key properties that need to be satisfied in flat multi-resource fair queueing~\cite{ghodsi2012multi}. Specifically, the fairness across flows can be guaranteed only when these properties are satisfied. In hierarchical scheduling, the fairness across sibling flows is also desired. Therefore, we modify these properties for hierarchical scheduling.

\vspace{0.3cm}
\noindent\textit{1) Hierarchical Share Guarantee}
\vspace{0.05in}

The essential property of fair queueing is \emph{isolation}. Traditionally, for a single resource (\emph{e.g.}, link bandwidth), fair queueing ensures that each of $N$ flows can get a guaranteed $\frac{1}{N}$ fraction of the bandwidth, regardless of the demand of other flows~\cite{srikant2013communication}.
Weighted fair queueing generalizes this concept by assigning each flow with a weight $\phi_i$, thus guaranteeing that flow $i$ can get at least $\frac{\phi_i}{\phi_q}$ of the bandwidth, where $q$ denotes a flat scheduler and $\phi_q = \sum_{j \in C(q)} \phi_j = \sum_{j\in C(P(i))} \phi_j$.
DRFQ further generalizes this from single resource to multiple resources such that any flow $i$ with weight $\phi_i$ receives at least $\frac{\phi_i}{\phi_q}$ fraction of one of the resources it uses~\cite{ghodsi2012multi}.
We then extend the guarantee of DRFQ from flat fair queueing to hierarchical fair queueing as follows.
\begin{defn}[Hierarchical Share Guarantee]
  A node in a hierarchy is supposed to receive at least $\frac{\phi_i}{\phi_{P(i)}}$ fraction of one of the resources it uses from its parent node.
\end{defn}

As a concrete example, consider two types of resources, CPU and link bandwidth, in the hierarchy shown in Fig.~\ref{fig_9}.
Assume there are $500$ time units available for the VNF scheduler $f_R$ to process the physical flows on each of the two resources.
Then both $f_1$ and $f_2$ should get at least $250$ time units on one of their resources, as they have the same weights. At one level down, the children of $f_1$ and $f_2$ should be guaranteed with time unites proportional to their weights from their parent.
That is, $f_{1,1}$ and $f_{2,1}$ each should get at least $100$ $(=250 \times 2/5)$ time units, while $f_{1,2}$ and $f_{2,2}$ each should get at least $150$ $(=250\times 3/5)$ time units on one of the two resources.

\vspace{0.3cm}
\noindent\textit{2) Strategy-Proofness}
\vspace{0.05in}

As discussed in DRFQ, the multi-resource setting is vulnerable to a new type of \emph{manipulation}. Flows can manipulate the scheduler to receive better service by purposely inflating their demand for resources they do not need.

The same kind of manipulations may also occur in hierarchical scheduling. Therefore, it is desired for multi-resource hierarchical schedulers to be resistant to them, as a system vulnerable to manipulations can incentivize users to waste resources, eventually leading to lower total goodput.

Strategy-proofness is the property that discourages the aforementioned manipulations, which is defined as follows,

\begin{defn}[Strategy-proofness]
  A flow should not be allowed to finish faster by increasing the amount of resources required to process it.
\end{defn}

\subsection{Review: Dominant Resource Fair Queueing} \label{sec:2.3}

DRFQ makes non-hierarchical scheduling decisions for packet processing.
Compared with the fair sharing principle in the single resource setting where each of the $n$ flows receives $1/n$ of the bandwidth, the allocation principle of DRFQ is for the flows to receive \textit{roughly} the same packet precessing time on their own dominant resources, that is, $W_i(t_1,t_2) = W_j(t_1,t_2)$ ($i\neq j$). This target can only be roughly achieved is due to the fact that a packet is indivisible and has to be scheduled as a whole. As a result, in practice, DRFQ is achieved by a greedy algorithm, progressive filling~\cite{bertsekas2021data}, that is always scheduling the next packet $p_i^k$ from the current poorest flow $i$, where the poorest flow $i$ is the one with the least $W_i$ at the moment. DRFQ continuously repeats this greedy scheduling algorithm, and then the order to forward packets from flows can be determined.

Specifically, according to the characteristics of scheduling requirements, the authors of~\cite{ghodsi2012multi} introduced a tradeoff between memoryless DRFQ and dove-tailing DRFQ, where memoryless and dove-tailing are two properties that cannot both achieved at the same time. We briefly illustrate the memoryless and dove-tailing DRFQ algorithms as follows.

\vspace{0.3cm}
\noindent\textit{1) Memoryless DRFQ}
\vspace{0.05in}

In memoryless DRFQ, a flow's current share of resources does not depend on its past share. Limiting memory is important to prevent flows from starvation~\cite{ghodsi2012multi}.
For example, if one flow uses a link at full rate for one minute, and a second flow becomes active, then with memory the second flow would occupy the link for the second minute to get a equal share with the first flow. Thus, the first flow starves for a minute.
Note that without memory, the dominant resource of a flow needs to be reconfirmed per-packet, and thus the memoryless DRFQ usually applies to flows where packets within the same flow have the same packet profiles.

Consider two flows in a middlebox sharing two types of resources $r_1$ and $r_2$, $f_1$ sends packets with profile $\langle 2,1\rangle$, $f_2$ sends packets with profile $\langle 1,2\rangle$. Then the dominant resource of $f_1$ is $r_1$ while the dominant resource of $f_2$ is $r_2$. Memoryless DRFQ greedily schedules the flow with the least current share and two flows will be scheduled alternatively such that the processing time of $f_1$ on $r_1$ and the processing time of $f_2$ on $r_2$ are the same.

\vspace{0.3cm}
\noindent\textit{2) Dove-tailing DRFQ}
\vspace{0.05in}

Dove-tailing DRFQ allows the scheduler to have memory of past processing time used by a flow.
This feature is referred as \emph{dove-tailing} in~\cite{ghodsi2012multi}. Since the scheduler records the processing time of past packets, dove-tailing DRFQ usually applies to flows where different packets from the same flow have different packet profiles.
For example, a flow that sends a total of 10 packets, alternating in processing time requirements $\langle 2,1\rangle$ and $\langle1,2\rangle$, respectively, is supposed to be treated the same as a flow that sends 5 packets, all with processing time $\langle 3,3 \rangle$.

It should be noted that although we assume that the packets within the same leaf have the same packet profiles, the packets experienced by an internal node may be different.
Therefore, modification of dove-tailing DRFQ is still necessary as it would be useful in our hierarchical fair queueing design when we consider the scheduling policy for internal nodes.

\subsection{Ineffectiveness of Naive Extensions}\label{sec:3}

Designing a hierarchical packet scheduler for multiple resources turns out to be non-trivial due to several problems that do not occur in flat scheduling.
We analyze two naive DRFQ extensions to support hierarchies and discuss their ineffectiveness.

\begin{figure}[t!]
\centering
  \subfloat[Simple example of a hierarchy.]{
    \includegraphics[width=1.7in,height=20mm]{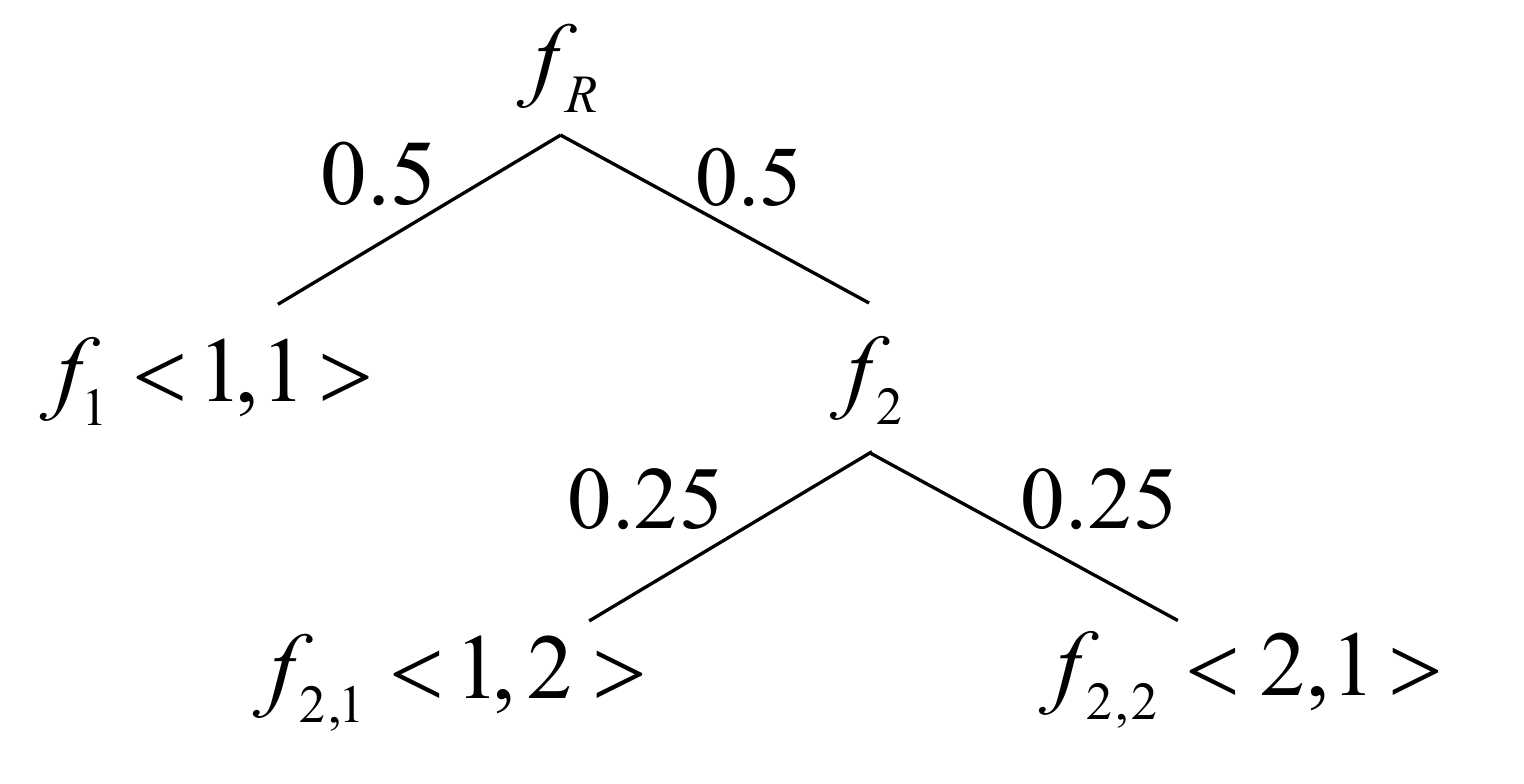}
    \label{fig_1:subfig:a}}
  \subfloat[First Attempt]{
    \includegraphics[width=1.5in,height=15mm]{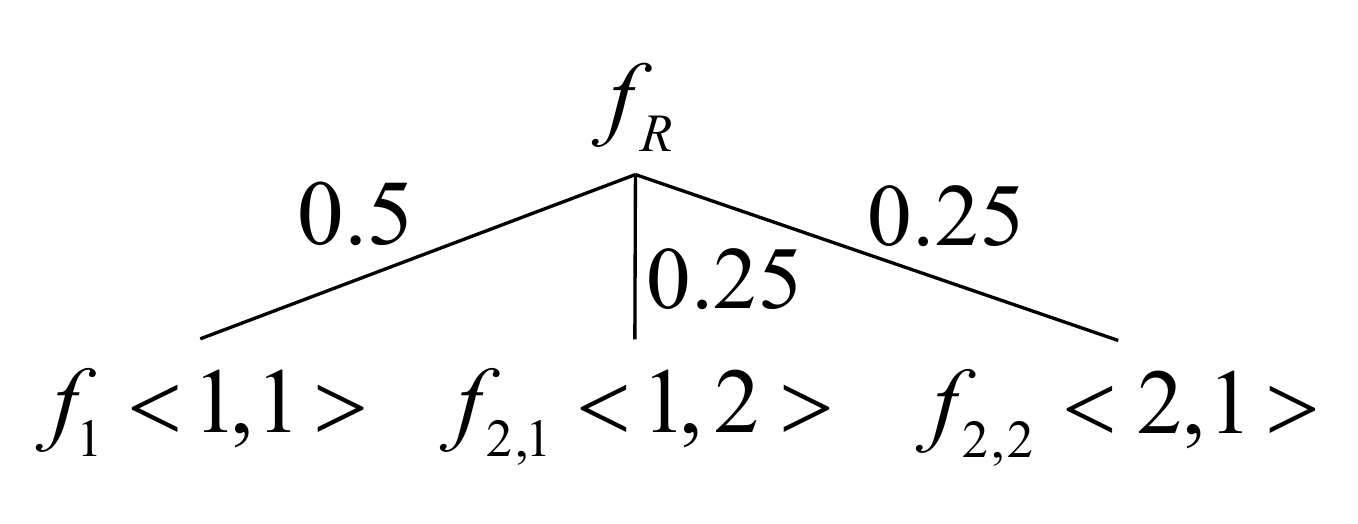}
    \label{fig_1:subfig:b}}
  \caption{Example of the naive extensions of memoryless DRFQ}
  \label{fig_1}
\end{figure}

\vspace{0.3cm}
\noindent\textit{1) Collapsed Hierarchies}
\vspace{0.05in}

The authors of~\cite{chandra2008hierarchical} proposed a well-known approach, collapsed hierarchies, which converts a hierarchical scheduler into a flat one. Its main idea is to take the hierarchy specification and compute the corresponding weight for each leaf node if the hierarchy is flattened. Although this method \emph{always} works well for a single resource~\cite{chandra2008hierarchical}, it turns out to be infeasible for multi-resource scheduling, as it violates the hierarchical share guarantee.

Consider the simple hierarchy shown in Fig.~\ref{fig_1:subfig:a}.
Our first attempt is to treat this hierarchy as the flat tree shown in Fig.~\ref{fig_1:subfig:b}. The weights of $f_{2,1}$ and $f_{2,2}$ keep the same as they both deserve half of the resources from their parent node $f_2$, which is also assigned half of the resources from the scheduler $f_R$. Meanwhile, the weight of $f_{1}$ is still 0.5. These weights accurately capture the share guarantee for each individual flow.
We now run the memoryless DRFQ to the three leaves shown in Fig.~\ref{fig_1:subfig:b} directly. The main idea is to equalize the weighted packet processing time across the three flows on their respective dominant resources. In this example, the dominant resources of $f_1$, $f_{2,1}$ and $f_{2,2}$ are $r_1$, $r_2$ and $r_1$, respectively.
Let $n_i$ denote the number of packets scheduled in $f_i$, then to equalized the weighted packet processing time on the dominant resources of $f_1$, $f_{2,1}$ and $f_{2,2}$, we have $\frac{n_1\times 1}{0.5} = \frac{n_{2,1}\times 2}{0.25} = \frac{n_{2,2}\times 2}{0.25}$, leading to a result such that $n_1:n_{2,1}:n_{2,2} = 4:1:1$.
Therefore, after an initial start, DRFQ obtains a periodic pattern in which the ratio of the number of packets scheduled in the three flows is $4:1:1$, making the length of the scheduling round period to be $7$ $(=4\times 1+1\times1+1\times2)$ time units. As a result, $f_{2,1}$ gets resource shares $\langle\frac{1}{7},\frac{2}{7}\rangle$ , while $f_{2,2}$ gets resource shares $\langle\frac{2}{7},\frac{1}{7}\rangle$, leading their parent $f_2$ to get resource shares $\langle\frac{3}{7},\frac{3}{7}\rangle$ . Therefore, the dominant resource share of $f_2$ is $\frac{3}{7}$, smaller than its guaranteed share 0.5, thus it violates the hierarchical share guarantee.

\vspace{0.3cm}
\noindent\textit{2) Naive Memoryless Extension}
\vspace{0.05in}

Inspired by the assumption that all the packets within the same leave have the same packet profiles, the second intuition is to apply memoryless DRFQ within each set of sibling nodes. For example for the hierarchy shown in Fig.~\ref{fig_1:subfig:a}, we first apply memoryless DRFQ to the pair of sibling nodes $f_{2,1}$ and $f_{2,2}$, and further apply memoryless DRFQ to the other pair of sibling nodes $f_1$ and $f_2$.
Surprisingly, this method also works well for a single resource while violating the hierarchical share guarantee in multi-resource scheduling.

Stick to the hierarchy shown in Fig.~\ref{fig_1:subfig:a}, after applying memoryless DRFQ to $f_{2,1}$ and $f_{2,2}$, the logical queue on node $f_2$ can be regarded as a flow alternating $\langle1, 2\rangle$ and $\langle2, 1\rangle$. We then apply memoryless DRFQ to $f_1$ and $f_2$.
Without memory, the dominant resource of $f_2$ is changing per-packet, and thus the scheduler will regard $f_2$ as a flow that is always requiring 2 units of packet processing time on its dominant resources (CPU or bandwidth).
To equalize the weighted packet processing time on the dominant resources of $f_1$ and $f_{2}$, we have $\frac{n_1\times 1}{0.5} = \frac{n_2 \times 2}{0.5}$. That is, every time one packet of $f_2$ with packet profile $\langle 1, 2\rangle$ or $\langle 2,1 \rangle$ is scheduled, two packets of $f_1$ would be scheduled accordingly. This leads to a result such that $n_1:n_{2,1}:n_{2,2} = 4:1:1$.
Therefore, after an initial start, DRFQ still obtains a periodic pattern in which the ratio of packet numbers scheduled in the three flows is $4:1:1$, making the length of the period to be $7$ $(=4\times1+1\times1+1\times2)$ time units.
Similar to the first intuitive method, here $f_2$ gets a dominant share of $\frac{3}{7}$, smaller than its guaranteed share 0.5. Therefore, the second attempt still fails in supporting hierarchical fair queueing.

The failure of naive memoryless DRFQ extensions to the hierarchical fair queueing necessitates alternative scheduling mechanisms, which are the main focus of the next section.

\section{Hierarchical Multi-resource Fair Queueing for Packet Processing}\label{sec:4}

To address the two issues discussed in the previous section, we propose two algorithms, collapsed H-DRFQ and dove-tailing H-DRFQ, which are presented in detail in the following two subsections, respectively.

\subsection{Collapsed Hierarchical Multi-resource Fair Queueing}\label{sec:4.1}

As we claim in the first naive extension in Section~\ref{sec:3}, our first method is to transform the hierarchy into a flat tree and then use the flat tree as an original memoryless DRFQ scheduler. However, as it was discussed in the collapsed hierarchies method in the previous section, directly inheriting the weights of leaves after the hierarchy is flattened violates the hierarchical share guarantee. Therefore, our challenge here is to compute the correct corresponding weight for each leaf node if the hierarchy is flattened. And the algorithm we propose based on the correct corresponding weights is named collapsed H-DRFQ.

In order to compute the correct weights for leaf nodes after the hierarchy is flattened, we first introduce a new concept, virtual packet profile, defined on internal nodes in the hierarchy. Given a hierarchy where all the leaves are backlogged, we can always regard an internal node as a virtual backlogged flow that submits packets with \emph{fixed} packet profiles. For example, in Fig.~\ref{fig_1:subfig:a}, $f_2$ can be regarded as a flow with resource requirement $\langle3,3\rangle$.
We refer the packet profile derived from this method as \emph{virtual packet profile}.
To compute the virtual packet profile in a general case, we introduce the normalized packet profiles computed by $l_{i,r}^k = \frac{s_{i,r}^k}{\mu_i^k}$, where $l_{i,r}^k$ denotes the normalized packet processing time of $p^k_{i}$ at resource $r$. Then the virtual packet profile of an internal node (\emph{i.e.}, an internal scheduler) can be formally defined as follows,

\begin{defn}[\textbf{Virtual Packet Profile}]\label{defn:3}
The virtual packet profile of a scheduler $q$ on packet $p_q^k$ is defined as the sum of the normalized packet processing time of all children of $q$ on their respective resources. That is,
\begin{equation}\label{equ:3.4}
  s_{q,r}^k = \sum_{i\in C(q)} l_{i,r}^k \phi_i.
\end{equation}
\end{defn}

The reason we use the normalized packet profile of leaf nodes to generate the virtual packet profile of internal nodes is that, in general, the weights of leaf nodes are not necessary to be the same, and the virtual packet profile of internal nodes should reflect the weighted contribution of each of its children. After normalization, the packet processing time of a leaf node on its dominant resource is always equal to 1 unit. At this very moment, adding weighted normalized packet profiles over all leaf nodes of scheduler $q$ can truly characterize its packet resource demand.

There is one more point to be noted. Given that all packets within the same leaf node are assumed to submit demands with the same packet profiles, \emph{i.e.}, $s_{i,r} = s_{i,r}^k$, each internal node will always keep a constant virtual packet profile during any backlogged period as long as its set of children does not change.

As an illustrative example, consider an internal node $\tilde{f}$ that has two children flow 1 and flow 2 with packet profiles $\langle 4, 2 \rangle$ and $\langle 1,2 \rangle$ sharing two resources. Thus the normalized packet profiles of flow 1 and flow 2 are $\langle \frac{4}{4}, \frac{2}{4} \rangle = \langle 1, 0.5 \rangle$ and $\langle \frac{1}{2},\frac{2}{2} \rangle = \langle 0.5, 1 \rangle$, respectively. If the weight of flow 1 is 0.4 and the weight of flow 2 is $0.2$, then based on Definition~\ref{defn:3}, the packet processing time of $\tilde{f}$ on $r_1$ is $1\times0.4+0.5\times0.2 = 0.5$, while the value on $r_2$ is $0.5\times0.4+1\times0.2 = 0.4$. Accordingly, the virtual packet profile of $\tilde{f}$ is $\langle 0.5,0.4 \rangle$.

\begin{figure}[t!]
\centering
  \subfloat[Minimum general sub-hierarchy.]{
    \includegraphics[width=1.5in, height=20mm]{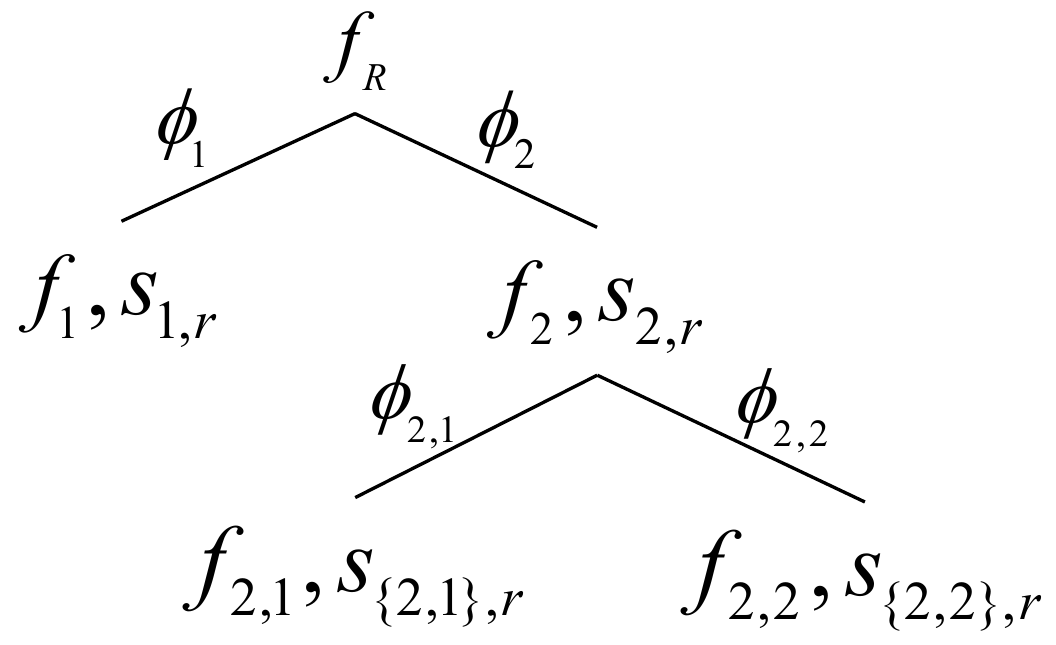}
    \label{fig_2:subfig:a}}
  \subfloat[Transformation of Fig. \ref{fig_2:subfig:a}.]{
    \includegraphics[width=1.7in, height=15mm]{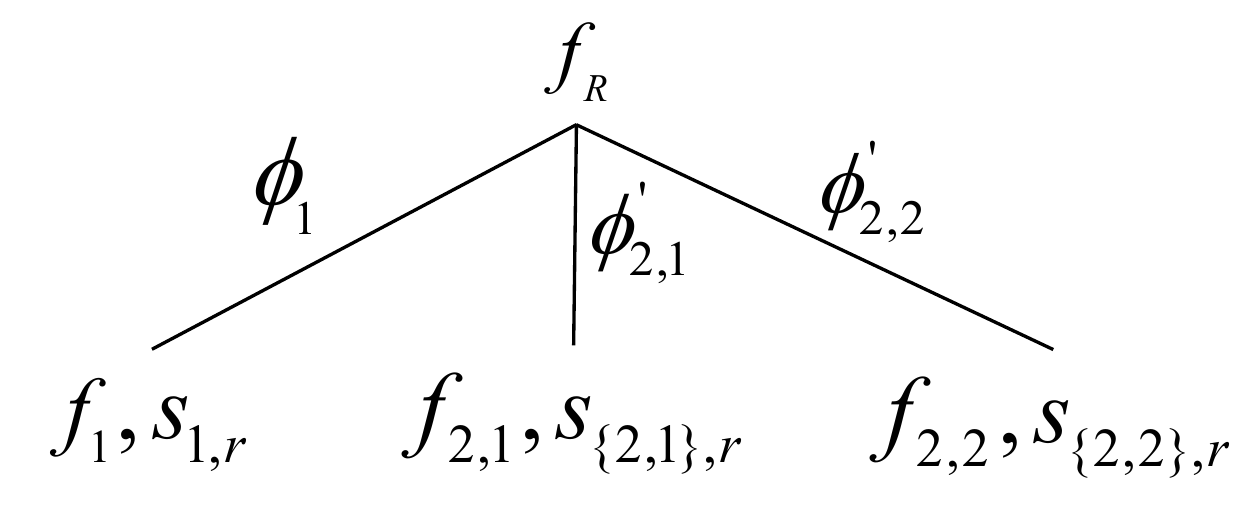}
    \label{fig_2:subfig:b}}
  \caption{Weight determination for the leaf node.}
  \label{fig_2}
\end{figure}

With the definition of virtual packet profile, we are now able to compute the weight for each leaf after converting the hierarchy into a flat one. To do so, we hope that the resources assigned to the corresponding leaves in the original hierarchy as well as the flattened tree are exactly the same. Specifically, let $n_i$ be the number of packets scheduled in $f_i$ from the hierarchy and $n'_i$ be the number of packets scheduled in the same node $f_i$ but from the flattened tree, it is desired for the collapsed H-DRFQ scheduler to achieve $n_i = n'_i$.

Consider a minimum general sub-hierarchy, which is able to form any hierarchical tree through iteration (see Fig.~\ref{fig_2:subfig:a}). Assume all the flows are backlogged throughout the example. Then achieving DRFQ allocation between each pair of sibling nodes in Fig.~\ref{fig_2:subfig:a} gives
\begin{subequations}\label{equ:3.5}
\begin{alignat}{3}
  \frac{n_{2,1}\mu_{2,1}}{\phi_{2,1}} & = \frac{n_{2,2} \mu_{2,2}}{\phi_{2,2}} \label{equ:3.5_1} \\
  \frac{n_1 \mu_1}{\phi_1} & = \frac{n_2 \mu_2}{\phi_2} \label{equ:3.5_2}\\
  n_2 s_{2,r} = n_{2,1} s_{\{2,1\},r} + & n_{2,2} s_{\{2,2\},r}, \quad (r=1,\dots,m) \label{equ:3.5_3}
\end{alignat}
\end{subequations}
Combining (\ref{equ:3.4}) and (\ref{equ:3.5_3}), we have
\begin{equation}\label{equ:3.6}
n_{2,1} = \frac{\phi_{2,1}}{\mu_{2,1}}n_2;n_{2,2} = \frac{\phi_{2,2}}{\mu_{2,2}}n_2.
\end{equation}

Meanwhile, to achieve DRFQ allocation between each pair of sibling nodes in Fig.~\ref{fig_2:subfig:b}, we have
\begin{equation}\label{equ:3.7}
    \frac{n'_1 \mu_1}{\phi_1} = \frac{n'_{2,1} \mu_{2,1}}{\phi'_{2,1}} = \frac{n'_{2,2} \mu_{2,2}}{\phi'_{2,2}}.
\end{equation}

In order to convert the hierarchy shown in Fig.~\ref{fig_2:subfig:a} into the flat one shown in Fig.~\ref{fig_2:subfig:b}, we have $n_1 = n'_1$, $n_{2,1} = n'_{2,1}$ and $n_{2,2} = n'_{2,2}$. Combining (\ref{equ:3.5_2}), (\ref{equ:3.6}) and (\ref{equ:3.7}), we have

\begin{equation}\label{equ:3.8}
\phi'_{2,1} = \frac{\phi_2\phi_{2,1}}{\mu_2};\phi'_{2,2} = \frac{\phi_2\phi_{2,2}}{\mu_2}.
\end{equation}

Therefore, in a general case where the hierarchy has an arbitrary tree topology, the weight for each leaf node in the flattened tree is
\begin{equation}\label{equ:3.9}
\phi'_i = \phi_i \prod_{h=1}^{H-1}\frac{\phi_{P^h(i)}}{\mu_{P^h(i)}}.
\end{equation}

Based on the formal discussion above, we present the pseudo-code of collapsed H-DRFQ in Algorithm~\ref{alg1}. When flows are added, removed, or change their demand status, Algorithm~\ref{alg1} is invoked to update the weights in the hierarchy.
\begin{algorithm}[h]\small
\caption{Collapsed H-DRFQ Pseudocode}
\label{alg1}

\DontPrintSemicolon
\SetKwFunction{FMain}{Collapsed H-DRFQ}
\SetKwProg{Fn}{Function}{:}{}
\Fn{\FMain{$f_i$}}{
    \For{each leaf node $f_i$}{
        Calculate $\phi'_i$\;
        Insert $f_i$ in the flat tree with weight $\phi'_i$ \;
        }
    Apply memoryless DRFQ to the flattened tree \;
}
\end{algorithm}
\begin{figure}[t!]
  \centering
  \subfloat[Example of a hierarchy]{
  \includegraphics[width=1.5in,height=20mm]{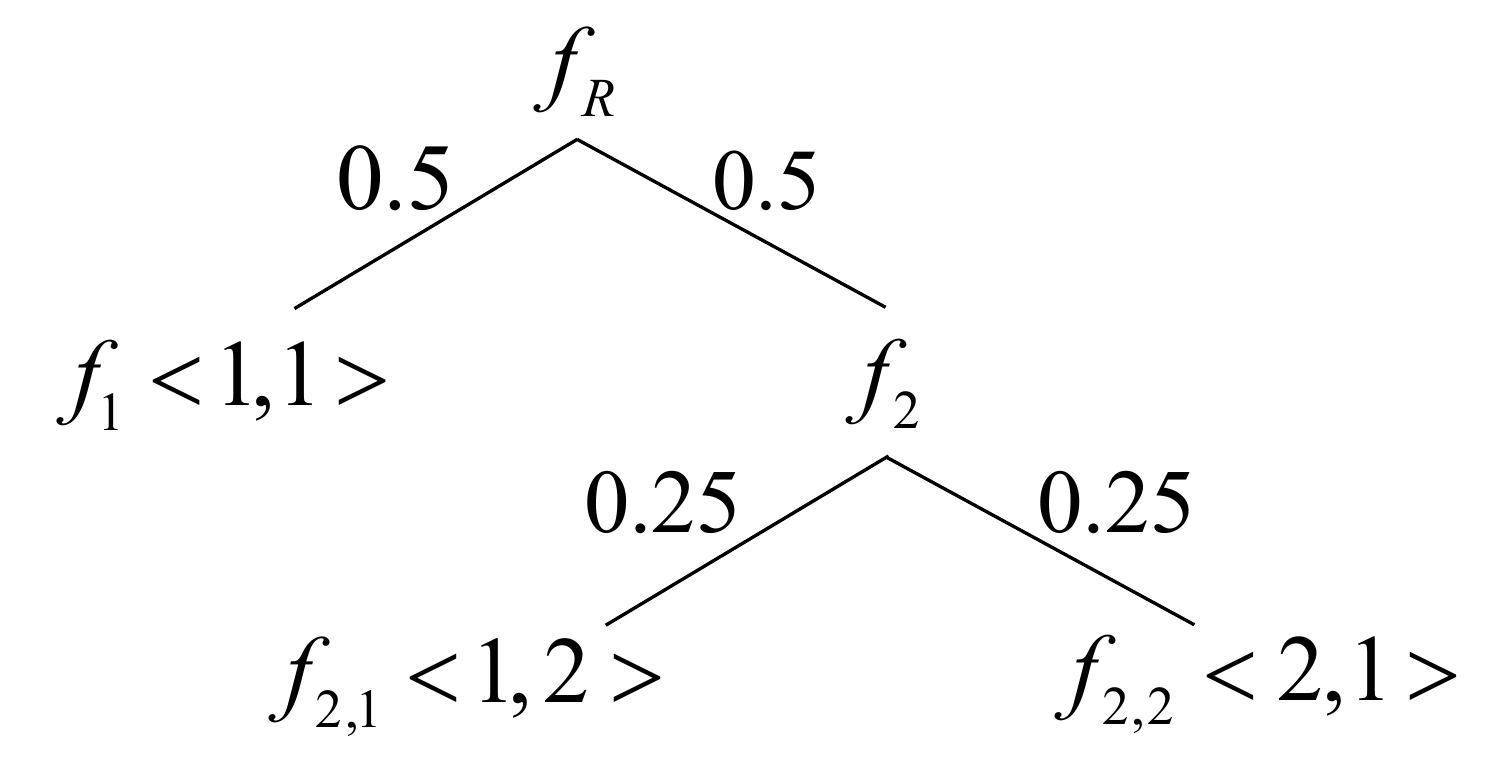}
  \label{fig_3:subfig:a}}
  \hfil
  \subfloat[Flattened tree of Fig. \ref{fig_3:subfig:a}]{
  \includegraphics[width=1.7in,height=15mm]{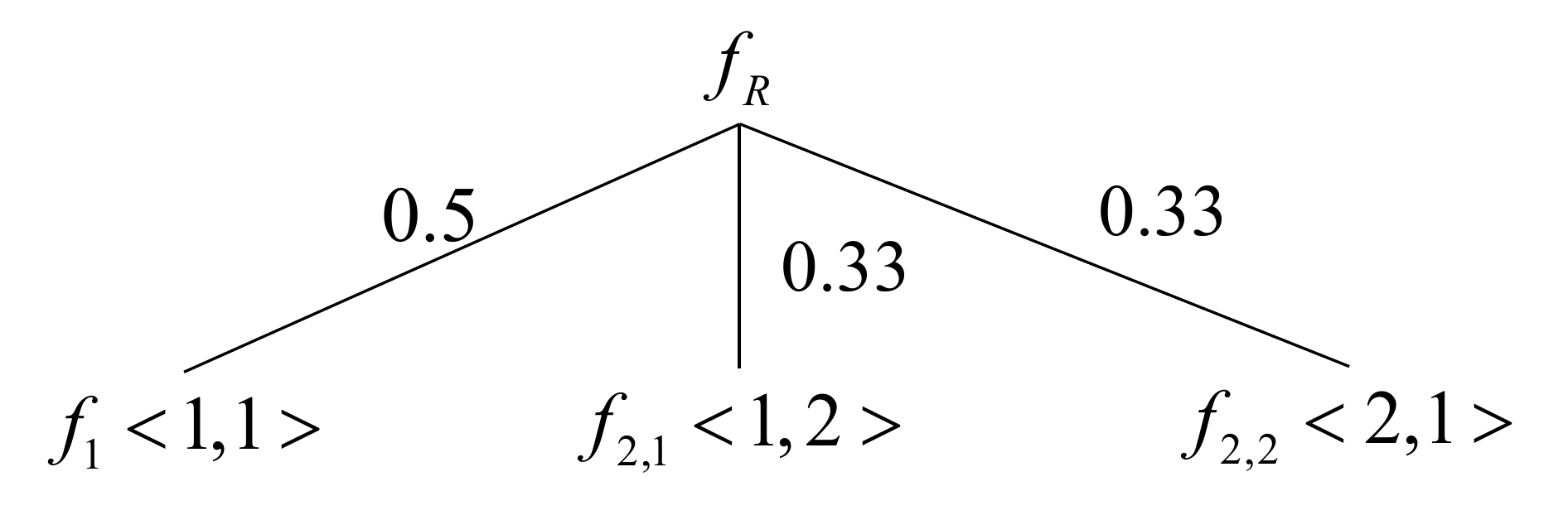}
  \label{fig_3:subfig:b}}
  \caption{Example hierarchy}
  \label{fig_3}
\end{figure}

Consider the hierarchy shown in Fig.~\ref{fig_3:subfig:a} where the three leaves $f_1$, $f_{2,1}$ and $f_{2,2}$ are assumed to be backlogged throughout the example. Two leaves $f_{2,1}$ and $f_{2,2}$ need to be flattened using collapsed H-DRFQ. We first compute the virtual packet profile of $f_2$. The normalized packet profile of $f_{2,1}$ and $f_{2,2}$ is $\langle 0.5,1\rangle$ and $\langle 0.5,1\rangle$ respectively. Then according to Equation~(\ref{equ:3.4}), the virtual packet profile of $f_2$ is $\langle (0.5+1)\times 0.25,(1+0.5)\times 0.25 \rangle = \langle 0.375, 0.375\rangle$. Therefore, we have $\mu_2 = 0.375$. Using Equation~(\ref{equ:3.9}), we have $\phi'_{2,1} = \phi_{2,1} \frac{\phi_{2}}{\mu_2} = \frac{0.25\times 0.5}{0.375} = \frac{1}{3} = 0.33$. Consequently, we obtain a flattened tree shown in Fig.~\ref{fig_3:subfig:b}. Then we use the tree in Fig.~\ref{fig_3:subfig:b} as a legacy weighted memoryless DRFQ scheduler to schedule the three physical flows $f_1$, $f_{2,1}$ and $f_{2,2}$. Based on the obtained weights after flattening, a memoryless DRFQ scheduler will schedule the three leaves such that $\frac{n_1}{\phi'_1} = \frac{2*n_{2,1}}{\phi'_{2,1}} = \frac{2*n_{2,2}}{\phi'_{2,2}}$. That is, $n_1:n_{2,1}:n_{2,2} = 3:1:1$. Meanwhile, greedily implementing collapsed H-DRFQ will always schedule the next packet from the poorest flow, leading to a flow scheduling order $f_1,f_{2,1},f_{2,2},f_1,f_1,f_1,f_{2,1},f_{2,2},f_1,f_1,f_1,,f_{2,1},f_{2,2}\ldots$ (see Fig.~\ref{fig_4:subfig:a}).

At last we analyse the time complexity of collapsed H-DRFQ algorithm.
In order to maintain consistency of complexity expression between the collapsed H-DRFQ and the dove-tailing H-DRFQ that we are going to discuss in the next section, we first introduce some notations for computing the structure of the hierarchy.
Although these notations seem to be unnecessary and complicated in computing the time complexity of collapsed H-DRFQ, it brings great convenience in expressing the time complexity of dove-tailing H-DRFQ.
Let $\Omega =\{\Omega_1,\dots,\Omega_T\}$ be the set of T sets of sibling nodes in a hierarchy. Take the hierarchy shown in Fig.~\ref{fig_9} for example, $T=4$ and $\Omega = \{ \{f_1,f_2\},\{f_{1,1}, f_{1,2}\},\{f_{2,1},f_{2,2}\}\}$. Let $|\Omega_i|$ denote the number of siblings in each set $\Omega_i \in \Omega$. Then we have Theorem~\ref{thm:5} showing the time complexity of collapsed H-DRFQ.

\begin{thm}\label{thm:5}
The time complexity of collapsed H-DRFQ is $O(\sum_{i=1}^{T} |\Omega_i|)$.
\end{thm}
\begin{IEEEproof}
Collapsed H-DRFQ runs in two stages. The first stage transforms the hierarchy into a flat tree. This process traverses the hierarchy to compute the weights for the leaves in the flattened tree and thus requires a $O(\sum_{i=1}^{T} |\Omega_i|)$ computing complexity.
The second stage is to schedule the flows based on the flattened tree with original DRFQ,
whose time complexity is $O(\log(B))$, the same with that of DRFQ, where $B$ denotes the number of leaves in the hierarchy. Therefore, the time complexity of collapsed H-DRFQ is $O(\sum_{i=1}^{T} |\Omega_i|+\log(B))=O(\sum_{i=1}^{T} |\Omega_i|)$.
\end{IEEEproof}

\subsection{Dove-tailing Hierarchical Multi-resource Fair Queueing}\label{sec:4.2}

Besides the collapsed H-DRFQ that first transforms a hierarchy into a flat tree, an alternative algorithm is to schedule sibling nodes directly in a DRFQ manner.
However, as it was discussed in the naive memoryless extension method in the previous section, applying memoryless DRFQ to each set of sibling nodes also leads to the violation of hierarchical share guarantees.
To address this problem, we propose the second H-DRFQ algorithm, named dove-tailing H-DRFQ, that uses the dove-tailing DRFQ method to each pair of sibling nodes in the hierarchy. Following this way, each internal node can be regarded as a logical queue that serves packets derived from its children, which is referred to as the \emph{logical packet profile}. Formally, we give the definition as follows.

\begin{defn}[\textbf{Logical Packet Profile}]\label{defn:4}
 At any given time $t$, the logical packet profile of a scheduler $q$ on packet $p_q^k$ is defined as packet profile of the packet $p_i^k$ of its child $f_i$ that is currently being scheduled by the scheduler. That is, $s_{q,r}^k(t) = s_{i,r}^k(t)$.
\end{defn}

Unlike the virtual packet profile that characterizes the resource demands using constant packet profiles, the logical packet profile of an internal is a combination of the packet profiles of its children, which may be changed over time.

Consider the same example shown in Fig.~\ref{fig_3:subfig:a}. Since the weight of $f_{2,1}$ and $f_{2,2}$ is the same, at any given time $t$, the logical packet profile of the internal node $f_2$ is $\langle1,2\rangle$ or $\langle2,1\rangle$. From a long-term perspective, $f_2$ acts as a logical flow alternating $\langle1,2\rangle$ and $\langle2,1\rangle$.

The logical packet profile provides another way to consider the packet processing time requirements of internal queues other than the virtual packet profile.
Packets traversing an internal node are from its children, leading the internal node to behave as a logical queue over time, with multiple packets that have different packet profiles.
Therefore, it is desirable for the scheduler to record the past packets served at each internal node. This is exactly the reason why dove-tailing H-DRFQ uses dove-tailing DRFQ between each pair of sibling nodes to provide hierarchical share guarantee.
The pseudo-code of dove-tailing H-DRFQ is summarized in Algorithm~\ref{alg2}.

\begin{algorithm}[h]\small
\caption{Dove-tailing H-DRFQ Pseudocode}
\label{alg2}
\DontPrintSemicolon
\SetKwFunction{FMain}{Dove-tailing H-DRFQ}
\SetKwProg{Fn}{Function}{:}{}
\Fn{\FMain{$f_i$}}{
    \If{$f_i$ is a leaf node}{
    \KwRet
    }
    \If{not each child of $f_i$ is visited}{
    Choose an unvisited child $f_{i'}$ of $f_i$ \;
    Run \texttt{Dove-tailing H-DRFQ}$(f_{i'})$ \;
    }
    Apply dove-tailing DRFQ to $f_i$ \;
    Enqueue the scheduling result to $f_i$'s parent's logical packet queue \;
    Mark $f_i$ as visited \;
}
\end{algorithm}

To demonstrate how does dove-tailing H-DRFQ work, we consider the same hierarchy shown in Fig.~\ref{fig_3:subfig:a} for flow scheduling. Implementing dove-tailing DRFQ to the children of $f_2$ makes its logical queue to be a flow with packet profiles $\langle1,2\rangle$ and $\langle2,1\rangle$ alternatively. Thereafter, the VNF scheduler implements dove-tailing DRFQ to $f_1$ and $f_2$, and the scheduler $f_R$ will obtain a logical queue, which is exactly the final order to schedule flows. That is, greedily implementing dove-tailing DRFQ to $f_1$ and $f_2$ will always schedule the next packet from the flow with the least current dominant share, leading to a flow scheduling order $f_1, f_{2,1}, f_1,f_1,f_{2,2},f_1,f_{2,1},f_1,f_1,f_{2,2}\ldots$ (see Fig.~\ref{fig_4:subfig:b}).

\begin{thm}\label{thm:6}
The time complexity of dove-tailing H-DRFQ is $O(\sum_{i=1}^T \log |\Omega_i|)$.
\end{thm}
\begin{IEEEproof}
Dove-tailing H-DRFQ actually implements dove-tailing DRFQ on each set of sibling nodes in a hierarchy. Therefore, the time complexity of dove-tailing H-DRFQ is the sum of the time complexity of dove-tailing DRFQ implementations on all the $T$ sets of sibling nodes. Consequently, the complexity of Dove-tailing H-DRFQ is $O(\sum_{i=1}^T \log |\Omega_i|)$.
\end{IEEEproof}

\section{Delay Analysis of H-DRFQ} \label{sec:5}

In this section, we examine the differences between the two H-DRFQ algorithms by comparing their delay bounds.

\subsection{A Simple Example Showing the Discrepancy}\label{sec:5.1}
We use the hierarchy in Fig.~\ref{fig_3:subfig:a} to coarsely illustrate the discrepancies between collapsed H-DRFQ and dove-tailing H-DRFQ. Assume all the three flows are backlogged throughout the example.

\begin{figure}[t!]
\centering
  \subfloat[The scheduling order of flows using collapsed H-DRFQ.]{
    \includegraphics[width=3in]{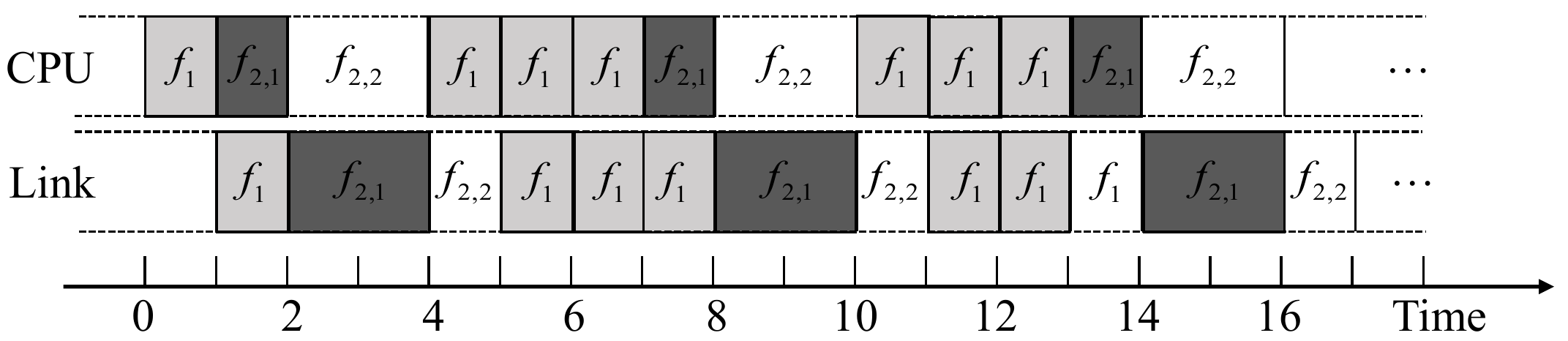}
    \label{fig_4:subfig:a}}
    \hfil
  \subfloat[The scheduling order of flows using dove-tailing H-DRFQ.]{
    \includegraphics[width=3in]{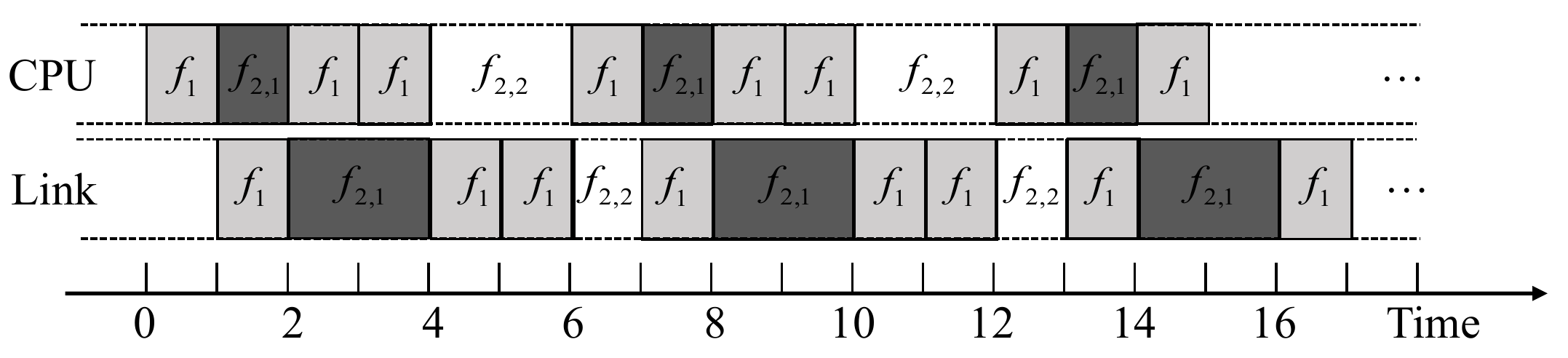}
    \label{fig_4:subfig:b}}
  \caption{A scheduler that implements H-DRFQ algorithms.}
  \label{fig_4}
\end{figure}

Fig.~\ref{fig_4} shows the scheduling order of the three physical flows using the two H-DRFQ algorithms.
After an initial start, the lengths of the periodic patterns of both algorithms are the same.
That is, in each periodic pattern, three packets from $f_1$, one packet from $f_{2,1}$ and one packet from $f_{2,2}$ are scheduled.
However, their packet order is different from each other. The scheduling order of flows using collapsed H-DRFQ is $f_{1},f_1,f_1,f_{2,1},f_{2,2}$ in each period, while the scheduling order of flows using dove-tailing H-DRFQ is $f_1,f_{2,1},f_1,f_1,f_{2,2}$ in each period.

The differences in scheduling orders may lead to different delays experienced by the same packet.
Therefore, we will analyse the scheduling delay to illustrate the differences between collapsed H-DRFQ and dove-tailing H-DRFQ in the following subsection.

\subsection{Multi-resource WFI} \label{sec:5.2}
To quantify the delay experienced by a packet, Worst-case Fair Index (WFI)~\cite{bennett1997hierarchical} is a common metric to characterize non-hierarchical fair queueing schedulers for a single resource (\emph{e.g.}, link sharing), which we quote as follows.

\begin{defn}[T-WFI]\label{defn:5}
A scheduler $q$ is said to guarantee a Time Worst-case Fair Index (T-WFI) of $\mathcal{A}_{i,q}$ for flow $i$, if for any time $t$, the delay of a packet arriving at $a_i^k$ is upper bounded by the following,
\begin{equation}\label{equ:4.1.a}
  d_i^k-a_i^k \leq \frac{Q_i(t)}{c_i} + \mathcal{A}_{i,q},
\end{equation}
where $a_i^k$ and $d_i^k$ are the arrival and departure times of $p_i^k$, respectively. $c_i$ is the rate guaranteed to flow $i$, which is $\frac{\phi_i}{\phi_q}$ of the total rate of the link $c_q$. $Q_i(t)$ is the number of bits in the queue at time $t$.
\end{defn}

Intuitively, $\mathcal{A}_{i,q}$ represents the maximum time a packet coming to an empty queue need to wait before receiving its guaranteed resource share. That is, $\mathcal{A}_{i,q}$ represents the delay bound of flow $i$. An important observation is that in an ideal fluid system in which the packets are infinitely divisible and multiple flows can receive service simultaneously, $\mathcal{A}_{i,q}$ is 0 since $f_i$ can receive its guaranteed share immediately after its arrival. However in a packet scheduling system, $\mathcal{A}_{i,q}$ must be larger than zero as a new arrival packet has to wait until the first arrivals finish in any one of the resources.

By multiplying $c_i$ on both sides of inequality (\ref{equ:4.1.a}), it is easy to transform the inequality into the form of following,
\begin{equation}\label{equ:4.1.b}
 W_i(a_i^k,d_i^k) \geq \frac{\phi_i}{\phi_q} W_q(a_i^k,d_i^k)-\alpha_{i,q},
\end{equation}
where $W_j(a_j^k,d_j^k)$ ($j=i$ or $j = q$) is the total amount of bits served by $j$ (individual flow $i$ or the scheduler $q$) during $[a_j^k,d_j^k]$, and $W_j(t_1,t_2) = c_j(t_2-t_1)$. $\frac{\phi_i}{\phi_q}$ is the resource share guaranteed to $f_i$. $\alpha_{i,q}$ represents a measurement of WFI in the unit of bits instead of seconds.

However, if we extend (\ref{equ:4.1.b}) to multiple resource settings, the ideal resource share received by $f_i$ may not be $\frac{\phi_i}{\phi_q}$. For instance, consider flow 1 and flow 2 with packet profiles $\langle1,2\rangle$ and $\langle2,1\rangle$ respectively, both flows have the same weights $0.5$. Every time $f_i$ is assigned $\langle \varepsilon, 2\varepsilon \rangle$, an ideal scheduler $q$ will then allocate $\langle 2\varepsilon, \varepsilon \rangle$ to $f_2$, making the processing time served by the scheduler to be $\langle3 \varepsilon, 3\varepsilon \rangle$. This defines an ideal fair queueing allocation with $W_1 (t_1,t_2) = W_2(t_1,t_2) = \frac{2}{3} \times W_q(t_1,t_2)$, where $W_i(t_1,t_2)$ denotes the packet processing time on the dominant resource of $f_i$. The ideal resource share $\frac{2}{3}$ is greater than the guaranteed resource share $\frac{0.5}{1} = \frac{1}{2}$. The problem in the multi-resource setting is that dove-tailing of resource demands, \emph{i.e}, that jobs have complementary packet profiles ($f_1$'s $\langle 1,2\rangle$ and $f_2$'s $\langle2,1\rangle$ in the example) gets more resource share.
We next propose a new theorem to rigorously quantify the ideal resource share $f_i$ received with DRFQ.
\begin{thm}\label{thm:3}
In DRFQ, the ideal resource share received by flow $i$ from its scheduler $q$ is
\begin{equation}\label{equ:4.2}
  \frac{W_i(t_1,t_2)}{W_q(t_1,t_2)} = \frac{\phi_i}{\mu_q},
\end{equation}
\end{thm}
\begin{IEEEproof}
According to the definition of virtual packet profile in Definition \ref{defn:3}, for any multi-resource scheduler $q$, we have $W_q(t_2,t_2) = \mu_q n_q$. Meanwhile, the processing time of flow $i$ on its dominant resource is $W_i(t_1,t_2) = \mu_i n_i$. According to (\ref{equ:3.6}), we have $n_i = \frac{\phi_i}{\mu_i} n_q$. Therefore, the ideal resource share received by flow $i$ from its scheduler $q$ should be
\begin{equation}\label{equ:4.2P}
  \frac{W_i(t_1,t_2)}{W_q(t_1,t_2)} = \frac{\mu_i n_i}{\mu_q n_q} = \frac{\mu_i \phi_i n_q}{\mu_i \mu_q n_q}=\frac{\phi_i}{\mu_q}.
\end{equation}\hspace{1em plus 1fill} \IEEEQEDhere
\end{IEEEproof}

At this point, we are able to introduce a new definition of WFI in Definition~\ref{defn:4} that is further applied to multi-resource packet processing.
\begin{defn}[M-WFI]\label{defn:6}
A scheduler $q$ is said to guarantee a Multi-resource Worst-case Fair Index (M-WFI) of $\alpha_{i,q}$, if for any interval $[t_1,t_2]$ within its backlogged period, the following holds
\begin{equation}\label{equ:4.3}
  W_i(a_i^k,d_i^k) \geq \frac{\phi_i}{\mu_q} W_q(a_i^k,d_i^k) -\alpha_{i,q},
\end{equation}
where $W_j(t_1,t_2)$ is the total processing time served by $f_j$ on its dominant resource during the time interval $[t_1,t_2]$. 
$\frac{\phi_i}{\mu_q}$ is the ideal resource share that $f_i$ is supposed to receive in DRFQ.
\end{defn}

In M-WFI, $\alpha_{i,q}$ represents the maximum processing time a packet $p_i^k$ coming to an idle flow $i$ needs to wait before receiving its deserved resource share. That is, $\alpha_{i,q}$ represents the processing time delay bound of flow $i$ in DRFQ.

\subsection{Delay Bounds of H-DRFQ}\label{sec:5.3}
DRFQ has already rigorously bounded the delay of a packet that arrives when a flow is idle~\cite{ghodsi2012multi}, which is shown in the next lemma. In this subsection, $s_{i,r}^\uparrow$ denotes $\max_k s_{i,r}^k$.

\begin{lem}\label{lemma:1}
Assume packet $p_i^k$ arrives at $t$, flow $i$ is idle at time $t$, and all packets have non-zero demand on every resource. Then the maximum delay to start serving packet $p_i^k$, $\alpha_{i,q}$, is bounded by
\begin{equation}\label{equ:4.4}
  \alpha_{i,q} \leq \max_r (\sum_{j\in Sib(i)} s_{j,r}^\uparrow).
\end{equation}
\end{lem}

We next bound the delay of a packet in H-DRFQ. The delay bounds of collapsed H-DRFQ and dove-tailing H-DRFQ are shown in Theorem~\ref{thm:1} and \ref{thm:2}, respectively. We follow the same assumption with~\cite{ghodsi2012multi}, \emph{i.e.}, each packet has a non-zero demand for every resource.

\begin{thm}\label{thm:1}
The collapsed H-DRFQ guarantees a delay bound $D^c(p_i^k)$ to start serving packet $p_i^k$
\begin{equation}\label{equ:4.5}
  D^c(p_i^k) \leq \max_r (\sum_{j \in Sib(i)} s_{j,r}).
\end{equation}
\end{thm}
\begin{IEEEproof}
In collapsed H-DRFQ, the scheduler firstly transforms the hierarchical structure into a flat tree and then schedules the flows according to DRFQ.
Therefore, the packet delay bound of collapsed H-DRFQ can be computed using the same method as DRFQ.
Meanwhile, as shown in~(\ref{equ:4.4}), for a flat scheduler, the packet delay bound has nothing to do with the flows' weights.
Therefore the packet delay bound of collapsed H-DRFQ is the same as that of DRFQ since collapsed H-DRFQ only changes the weights of physical flows after flatting the hierarchy.
Moreover, the packet profiles within the same leaf are assumed to be the same in collapsed H-DRFQ, \emph{i.e.}, $s_{i,r}^\uparrow = s_{i,r}$. Thus the packet delay bound is $\max_r (\sum_{j \in Sib(i)} s_{j,r})$. \hspace{1em plus 1fill} \IEEEQEDhere
\end{IEEEproof}

\begin{thm}\label{thm:2}
The dove-tailing H-DRFQ guarantees a delay bound $D^t(p^k_i)$ to start serving packet $p_i^k$
\begin{equation}\label{equ:4.6}
\begin{split}
   D^t(p_i^k) \leq & \sum_{w=1}^{H-1}\prod_{h=0}^{w-1} \max_r (\sum_{j\in\mathcal{U}(Sib(P^w(i)))}\frac{\phi_{P^h(i)}}{\mu_{P^{h+1}(i)}} s_{j,r})+ \\
     &  \max_r(\sum_{j\in Sib(i)} s_{j,r}),
\end{split}
\end{equation}
where $\mathcal{U}(i)$ is the set of leaves sharing the same ancestor $f_i$, and $\mathcal{U}(Sib(P^w(i)))$ is the set of leaves sharing the same set of ancestors which are sibling nodes of $P^w(i)$.
\end{thm}
\begin{IEEEproof}
Since node $P^{h+1}(i)$ is the worst-case fair with the logical queue at node $P^h(i)$, the following holds for
\begin{equation}\label{equ:4.7}
  W_{P^h(i)}(a_i^k,d_i^k) \geq \frac{\phi_{P^h(i)}}{\mu_{P^{h+1}(i)}} W_{P^{h+1}(i)}(a_i^k,d_i^k) - \alpha_{P^h(i)},
\end{equation}
where $W_{P^h(i)}(a_i^k,d_i^k)$ is the amount of service received by flow $P^h(i)$ in $[a_i^k,d_i^k]$, and $h = 0,\dots,H-1$. Then we have

\begin{equation}\label{equ:4.9}
\begin{split}
 W_i(a_i^k,d_i^k) \geq & \frac{\phi_i}{\mu_{P(i)}} W_{P(i)}(a_i^k,d_i^k) - \alpha_i \\
 \geq &\frac{\phi_i}{\mu_{P(i)}}[ \frac{\phi_{P(i)}}{\mu_{P^2(i)}} W_{P^2(i)}(a_i^k,d_i^k) - \alpha_{P(i)}]-\alpha_i \\
  = &\frac{\phi_i \phi_{P(i)}}{\mu_{P(i)} \mu_{P^2(i)}} W_{P^2(i)}(a_i^k,d_i^k) - \frac{\phi_i}{\mu_{P(i)}} \alpha_{P(i)} - \alpha_i \\
  \geq &\dots \\
  \geq & \frac{\phi_i \phi_{P(i)} \dots \phi_{P^{H-1}(i)}}{\mu_{P(i)}\mu_{P^2(i)} \dots \mu_{P^H(i)}} W_{P^H(i)}(a_i^k,d_i^k) -\\
  & (\frac{\phi_i \phi_{P(i)} \dots \phi_{P^{H-2}(i)}}{\mu_{P(i)}\mu_{P^2(i)}\dots \mu_{P^{H-1}(i)}}\alpha_{p^{H-1}(i)} + \\
  & \dots + \frac{\phi_i}{\mu_{P(i)}}\alpha_{P(i)} + \alpha_i) \\
   \geq & \prod_{h=0}^{H-1} \frac{\phi_{P^h(i)}}{\mu_{P^{h+1}(i)}} W_R(a_i^k,d_i^k) -\\
  & (\sum_{w=1}^{H-1} \prod_{h=0}^{w-1} \frac{\phi_{P^h(i)}}{\mu_{P^{h+1}(i)}} \alpha_{P^w(i)}+\alpha_i),
\end{split}
\end{equation}
where $W_R(t_1,d_i^k) = W_{P^H(i)}(a_i^k,d_i^k)$ denotes the total processing time consumed by the root node $R$ on its dominant resource.
Therefore, we have
\begin{equation}\label{equ:4.10}
  D^t(p_i^k) = \sum_{w=1}^{H-1} \prod_{h=0}^{w-1} \frac{\phi_{P^h(i)}}{\mu_{P^{h+1}(i)}} \alpha_{P^w(i)}+\alpha_i.
\end{equation}
Meanwhile, Lemma~\ref{lemma:1} indicates that
\begin{equation}\label{equ:4.11}
  \alpha_{P^h(i)} \leq \max_r (\sum_{j\in Sib(p^h(i))} s_{j,r}^\uparrow).
\end{equation}
For each internal node $f_{P^h(i)}$ ($0<h<H$), we have
\begin{equation}\label{equ:4.15}
  s_{P^h(i),r}^\uparrow = \max_{j\in C(P^h(i))}(s_{j,r}^\uparrow ) \leq \sum_{j\in C(P^h(i))} s_{j,r}^\uparrow,
\end{equation}
where $s_{P^h(i),r}^\uparrow$ denotes the maximal packet processing time requirement of the internal logical flow $f_{P^h(i)}$ on resource $r$. The first step uses the fact that $s_{P^h(i),r}^\uparrow$ equals to one of its children that has the maximum $s_{j,r}^\uparrow$ for $j \in C(P^h(i))$, and the second step uses the fact that the maximum $s_{j,r}^\uparrow$ among $f_{P^h(i)}$'s children must be less than or equal to the sum of $s_{j,r}^\uparrow$ of $f_{P^h(i)}$'s children.

Combining (\ref{equ:4.10}), (\ref{equ:4.11}) and (\ref{equ:4.15}) yields
\begin{equation}\label{equ:4.16}
\begin{split}
    \alpha_{P^w(i)} \leq & \max_r (\sum_{j \in Sib(P^w(i))} s_{j,r}^\uparrow) \\
     \leq & \max_r (\sum_{j \in Sib(P^w(i))} \sum_{v \in C(j)} s_{v,r}^\uparrow)\\
     \leq & \max_r (\sum_{j \in Sib(P^w(i))} \sum_{v \in C(j)} \sum_{\theta \in C(v)} s_{\theta,r}^\uparrow)\\
     \leq & \dots \\
     \leq & \max_r (\sum_{j \in \mathcal{U}(Sib(P^w(i)))} s_{j,r}^\uparrow)
\end{split}
\end{equation}
Moreover, notice that (\ref{equ:4.10}) can be regarded as a linear mapping from $\alpha_{P^w(i)}$ ($w=0,\dots,H-1$) to $D^t(P^k_i)$, where $\prod_{h=0}^{w-1} \frac{\phi_{P^h(i)}}{\mu_{P^{h+1}(i)}}$'s are coefficients (also viewed as weights) parameterizing the space of linear functions. Consequently, using (\ref{equ:4.16}), the right-hand-side of (\ref{equ:4.10}) can be further bounded as flowing,
\begin{equation}\label{equ:4.12}
\begin{split}
D^t(p_i^k) \leq  &\sum_{w=1}^{H-1}\prod_{h=0}^{w-1} \max_r(\sum_{j\in \mathcal{U}(Sib(P^w(i)))} \frac{\phi_{P^h(i)}}{\mu_{P^{h+1}(i)}} s_{j,r})+\\
    & \max_r(\sum_{j\in Sib(i)} s_{j,r}).
\end{split}
\end{equation}
Here we also remove the uparrow of $s_{j,r}$ since in (\ref{equ:4.12}) $s_{j,r}$ denotes the packet processing time requirements of a physical flow $j$ on resource $r$, which are assumed to be the same within the same physical flow.
\hspace{1em plus 1fill} \IEEEQEDhere
\end{IEEEproof}

The next corollary compares the delay bounds exhibited by collapsed H-DRFQ and dove-tailing H-DRFQ respectively.

\begin{cor}\label{cor:2}
    The delay bound of dove-tailing H-DRFQ $D^t(p_i^k)$ is always smaller than that of collapsed H-DRFQ $D^c(p_i^k)$. That is,
    \begin{equation}\label{equ:4.14}
      \max(D^t(p_i^k)) \leq \max(D^c(p_i^k)).
    \end{equation}
\end{cor}
\begin{IEEEproof}
  We observe that the upper bound of $D^t(p_i^k)$ (see (\ref{equ:4.6})) can be regarded as a linear function of $\max_r (\sum_{j\in \mathcal{U}(Sib(P^w(i)))} s_{j,r})$ ($w=0,\dots,H-1$), with each term parameterized by $\prod_{h=0}^{w-1} \frac{\phi_{P^h(i)}}{\mu_{P^{h+1}(i)}}$. Meanwhile, the upper bound of $D^c(p_i^k)$ (see (\ref{equ:4.5})) can be also regarded as a linear function of $\max_r (\sum_{j\in \mathcal{U}(Sib(P^w(i)))} s_{j,r})$ ($w=0,\dots,H-1$), with each term parameterized by 1. Consequently, the upper bound of $D^t(p_i^k)$ is a linear transformation of the upper bound of $D^c(p_i^k)$. Moreover, $\prod_{h=0}^{w-1} \frac{\phi_{P^h(i)}}{\mu_{P^{h+1}(i)}}$ is smaller than or equal to $1$, as $\frac{\phi_{p^h(i)}}{\mu_{P^{h+1}(i)}}$ denotes the ideal dominant share of $f_{P^h(i)}$ and is always less than or equal to 1. Therefore, we have $\max(D^t(p_i^k)) \leq \max(D^c(p_i^k))$.\hspace{1em plus 1fill} \IEEEQEDhere
\end{IEEEproof}

\subsection{Discussion\label{sec:6.4}}
Although the delay bound of dove-tailing H-DRFQ is always smaller than or equal to that of collapsed H-DRFQ (\emph{i.e.}, $ \max(D^t(p_i^k)) \leq \max(D^c(p_i^k))$), this does not necessarily indicate that dove-tailing H-DRFQ always outperforms collapsed H-DRFQ in terms of packet delays. We will verify this in Section \ref{sec:5.2} through extensive simulations. In this subsection, we only give qualitative analysis.

The delay bound of $D^c(p_i^k)$ may be not tight.
Intuitively, a packet comes to an idle flow may have to wait for a long time.
This is because some packets related to that idle flow have received \emph{more} service than expected in a previous time period. In the case of non-hierarchical fair queueing, these packets must belong to the same flow $i$. In the case of hierarchical fair queueing, these packets may belong to the flows that share the same ancestors with flow $i$.
Meanwhile, WFI does not bound delay tightly using a flat or non-hierarchical scheduler since it does not take into account the fact that packets from the same flow may receive more service in a previous time period. However, WFI is more important in bounding the delay using a hierarchical scheduler because the extra service received in the previous time period may have been received by a flow other than the one being considered~\cite{bennett1997hierarchical}.


\section{H-DRFQ Properties}\label{sec:6}

We have shown that naive extensions of DRFQ for hierarchies failed in supporting certain guarantees and thus propose collapsed H-DRFQ and dove-tailing H-DRFQ to address this problem. In this section we explore important properties of H-DRFQ.

\subsection{Hierarchical Share Guarantee}\label{sec:6.1}
\begin{cor}\label{cor:1}
The ideal dominant share received by physical flow $f_i$ in H-DRFQ is
\begin{equation}\label{equ:4.13}
  \prod_{h=0}^{H-1}\frac{\phi_{P^h(i)}}{\mu_{P^{h+1}(i)}}.
\end{equation}
\end{cor}
\begin{IEEEproof}
On the one hand, in collapsed H-DRFQ, the weight of each leaf in the flattened tree is shown in (\ref{equ:3.9}). Then the ideal dominant share received by each leaf in the hierarchy equals to the dominant share receive by each leaf in the flattened tree. According to Theorem~\ref{thm:3}, the ideal dominant share received by each physical flow $i$ in the flattened tree can be computed as,
\begin{equation}\label{equ:4.17}
\begin{split}
   \frac{W_i(a_i^k, d_i^k)}{W_R(a_i^k, d_i^k)} & = \frac{\phi'_i}{\mu_{P^H(i)}} \\
     & = \frac{\phi_i }{\mu_{P^H(i)}} \prod_{h=1}^{H-1}\frac{\phi_{P^h(i)}}{\mu_{P^h(i)}}\\
     &  = \frac{\phi_i \phi_{P(i)}\dots \phi_{P^{H-1}(i)}}{\mu_{P(i)} \mu_{P^2(i)} \dots \mu_{P^H(i)}} \\
     & = \prod_{h=0}^{H-1}\frac{\phi_{P^h(i)}}{\mu_{P^{h+1}(i)}}.
\end{split}
\end{equation}

On the other hand, in dove-tailing H-DRFQ, the WFI of the VNF scheduler is given in (\ref{equ:4.9}). It is obvious that the ideal dominant share received by $f_i$ is the multiplier of $W_R(a_i^k,d_i^k)$, which is $\prod_{h=0}^{H-1}\frac{\phi_{P^h(i)}}{\mu_{P^{h+1}(i)}}$.
\end{IEEEproof}

Given this ideal resource share received by a flow in a hierarchy, we can rigorously prove that H-DRFQ can provide hierarchical share guarantees.

\begin{thm}[\textbf{Hierarchical Share Guarantee}]\label{thm:4}
H-DRFQ satisfies the hierarchical share guarantee property.
\end{thm}
\begin{IEEEproof}
According to the method of generating virtual packet profile of internal queues shown in (\ref{equ:3.4}), we have
\begin{equation}\label{equ:5.1}
   s_{q,r}^k  = \sum_{i\in C(q)} l_{i,r}^k \phi_i = \sum_{i\in C(q)} \frac{s_{i,r}^k}{\mu^k}\phi_i \leq \sum_{i\in C(q)} \phi_i = \phi_q.
\end{equation}
Thus, $\mu_q = \mu_q^k = \max_r(s_{q,r}^k) \leq \phi_q$. Therefore, for each item in (\ref{equ:4.13}), we have
\begin{equation}\label{equ:5.2}
  \frac{\phi_{P^h(i)}}{\mu_{P^{h+1}(i)}} \geq \frac{\phi_{P^h(i)}}{\phi_{P^{h+1}(i)}},
\end{equation}
as $\mu_{P^{h+1}(i)} \leq \phi_{P^{h+1}(i)}$ for $h = 0,\dots, H-1$. Since $\frac{\phi_{P^h(i)}}{\phi_{P^{h+1}(i)}}$ represents the hierarchical share guarantee we defined in Section~\ref{sec:2.2}, (\ref{equ:5.2}) indicates that H-DRFQ satisfies hierarchical share guarantee property.\hspace{1em plus 1fill} \IEEEQEDhere
\end{IEEEproof}

\subsection{Group Strategy-Proofness}\label{sec:6.2}



In the context of hierarchical multi-resource fair queueing, the flows within a group could collude to manipulate schedulers. To address this issue, it is desirable that the allocation policy satisfies group strategy-proofness, a hierarchical extension of strategy-proofness in DRFQ.

\begin{thm}[Group Strategy-proofness]
  H-DRFQ satisfies the group strategy-proofness property.
\end{thm}

\begin{IEEEproof}
  Assume a flow $f_i$ is able to finish faster by increasing the amount of resources allocated to it from $s_{i,r}$ to $s'_{i,r}$. That is, $s'_{i,r} > s_{i,r}$. Assume that the number of packets being processed in $f_i$ in unit time is $\tilde{n}_i$, and the number of packets being processed in $f_i$ in unit time after $f_i$ increases its resource demands is $\tilde{n}'_i$. Since we assume that $f_i$ is able to finish faster by increasing its resource demands, we have $\tilde{n}'_i > \tilde{n}_i$. Originally, in order to equalize the dominant share between each pair of sibling nodes, the dominant share of $f_i$ should be equal to the dominant share of its sibling node $f_j$. That is, $s_{i,r}\tilde{n}_i = s_{j,r}\tilde{n}_j$. After $f_i$ increases its resource demands, the dominant shares of $f_i$ and $f_j$ should still be the same, that is, $s'_{i,r}\tilde{n}'_i = s_{j,r}\tilde{n}_j$. As $s'_{i,r} > s_{i,r}$, we should have $\tilde{n}'_i < \tilde{n}_i$. This obviously contradicts with the assumption that $\tilde{n}'_i > \tilde{n}_i$. Therefore, a flow should not be able to finish faster by increasing its resource demands. In conclusion, H-DRFQ satisfies the strategy-proofness property.
\end{IEEEproof}

\section{Implementation}\label{sec:7}

\begin{table}[!t]
\centering
\caption{Linear model for CPU processing time in 3 middlebox modules~\cite{ghodsi2012multi}. Model parameters are based on the measurement results.\label{tab:4}}
{\scriptsize
\begin{tabular}{|c|c|}
  \hline
  \textbf{Module} & \textbf{CPU processing time $(\mu s)$} \\ \hline
  \textbf{Basic Forwarding} & 0.00286 $\times$ PacketSizeInBytes + 6.2 \\ \hline
  \textbf{Statistical Monitoring} & 0.0008 $\times$ PacketSizeInBytes + 12.1 \\ \hline
  \textbf{IPSec Encryption} & 0.015 $\times$ PacketSizeInBytes + 84.5\\ \hline
\end{tabular}}
\end{table}

To verify the properties and demonstrate H-DRFQ algorithms, we implement them in a testbed based on Click modular router~\cite{kohler2000click}.
%
%
Specifically, we implement the simple hierarchy shown in Fig.~\ref{fig_2:subfig:a}, where the weight of the three physical flows $f_1$, $f_{2,1}$ and $f_{2,2}$ is set to be $0.5$, $0.25$ and $0.25$, respectively.

\subsection{Configurations of the Testbed} \label{sec:7.1}

\begin{figure}[t!]
  \centering
  \subfloat[The collapsed H-DRFQ configuration]{
  \includegraphics[width=1.7in]{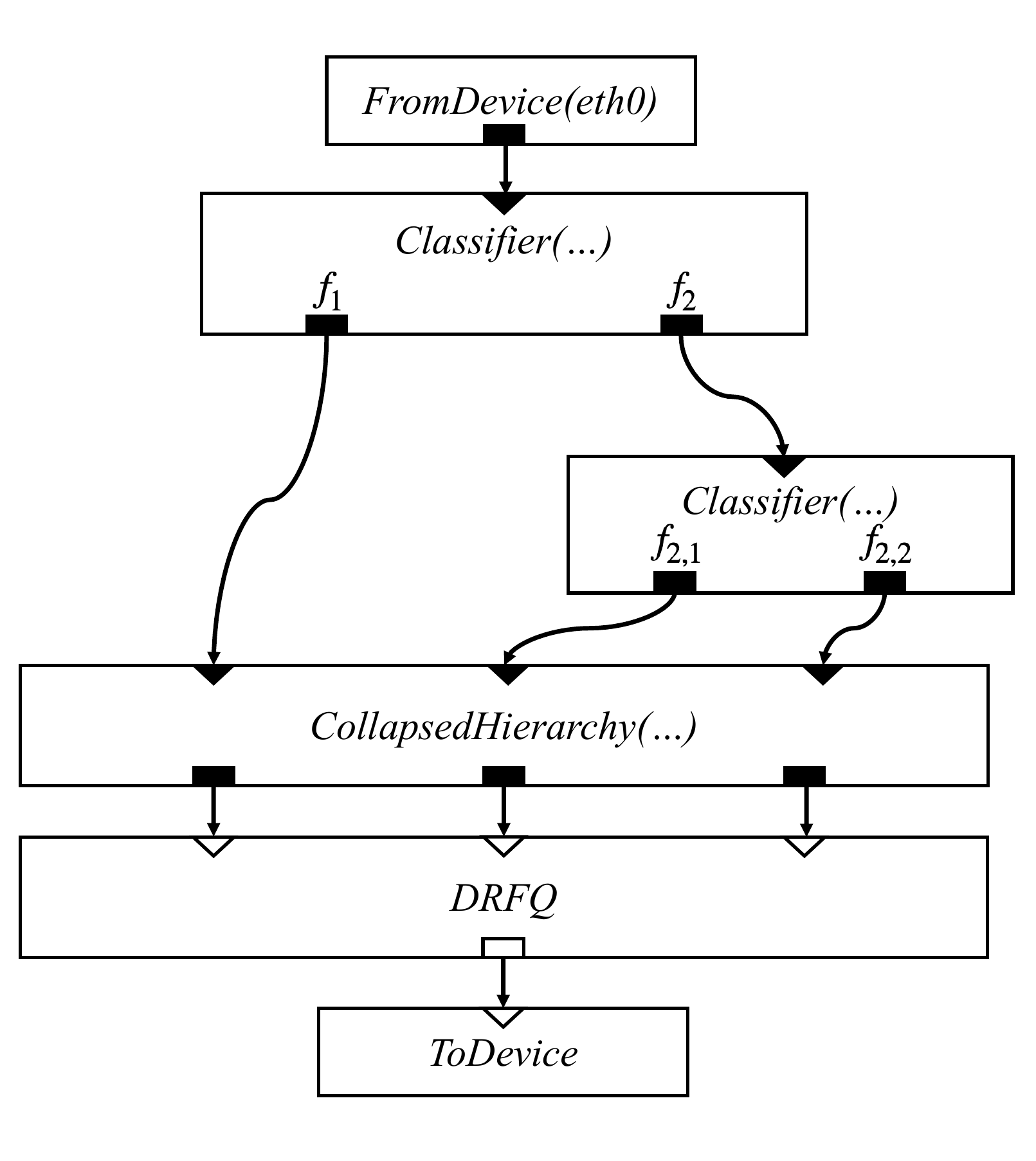}
  \label{fig_implementation:subfig:a}}
  \hfil
  \subfloat[The dove-tailing H-DRFQ configuration]{
  \includegraphics[width=1.5in]{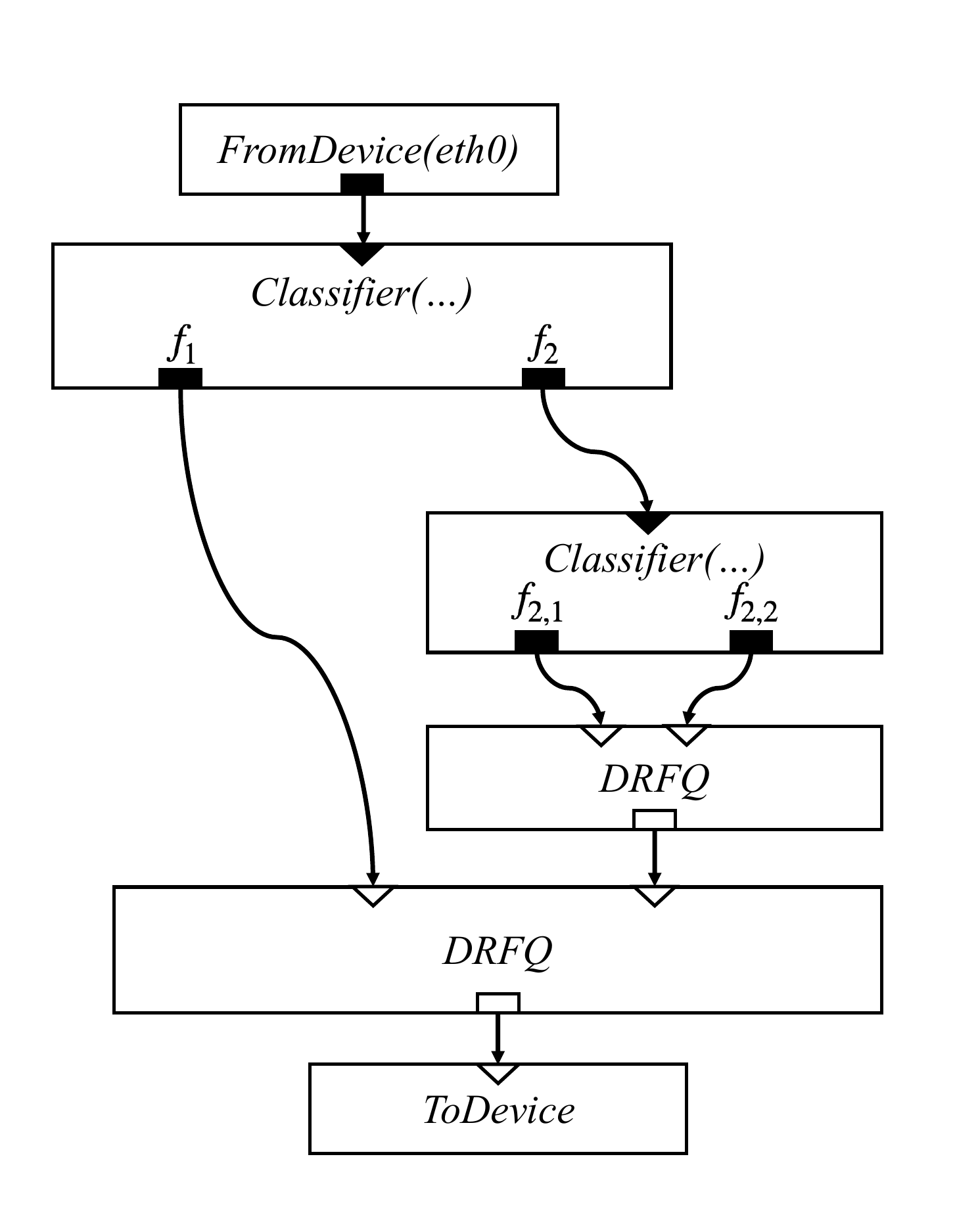}
  \label{fig_implementation:subfig:b}}
  \caption{The H-DRFQ configuration}
  \label{fig_implementation}
\end{figure}

The configurations of the testbed for implementing the two H-DRFQ algorithms are shown in Fig.~\ref{fig_implementation}.
We first re-implement the DRFQ module presented in~\cite{ghodsi2012multi}, which is also a part of our H-DRFQ implementation.
Then, for collapsed H-DRFQ, as it is shown in Fig.~\ref{fig_implementation:subfig:a}, the \emph{Classifier} module will first classify the flows into two service classes according to their packets' differentiated services code point (DSCP), then, for flow group $f_2$, the \emph{Classifier} module will further classify it into two flows $f_{2,1}$ and $f_{2,2}$ according to their IP addresses. To implement H-DRFQ we add a new \emph{CollapsedHierarchy} module to Click. The \emph{CollapsedHierarchy} module is designed for identifying the structure of the incoming flows based on port numbers and IP prefixes (including IP address and DSCP). It computes the weights of individual flows after the hierarchy is flattened and then forwards packets to each corresponding output. Then, the DRFQ module will schedule the three flows as a flat scheduler.

For implementing dove-tailing H-DRFQ, as shown in Fig.~\ref{fig_implementation:subfig:b}, no more modules would be added except for the DRFQ module. The incoming flows will be similarly classified by the \emph{Classifier} module. But at this time, $f_{2,1}$ and $f_{2,2}$ will be first scheduled by the DRFQ module and an integrated flow is output, which is taken as an input to the next DRFQ module along with another input $f_1$.

We run a Click-based multi-function middlebox in usermode on a machine with a 3.1 GHz Intel ``Core i5" processor (7267U) and an 1 Gbps Ethernet link. We connect this machine to a computer that uses the \emph{FastUDPSource} module in Click router to send packets. Since our machine only has an 1 Gbps link, we throttled its outgoing bandwidth to 200 Mbps to emulate a congested link, and the fraction of CPU time that the DRFQ module is allowed to use is set to be 20\% so that the CPU can also be a bottleneck at this rate.

We begin by generating three flows using the \emph{FastUDPSource} module, each sending 25,000 1300-byte UDP packets per second to exceed the total outgoing bandwidth capacity. We configure the flows such that: (1) $f_1$ only needs basic forwarding, which requires bandwidth the most, (2) $f_{2,1}$ is statistical monitoring flow, which also requires bandwidth the most but uses slightly more CPU than basic forwarding, (3) $f_{2,2}$ is an IPSec flow, which requires CPU the most. According to~\cite{ghodsi2012multi}, the estimate of CPU processing time linearly increases with the packet size and such linear formulations are summarized in Table~\ref{tab:4}.

\subsection{Dynamic Allocation Results} \label{sec:7.2}

\begin{figure}[!t]
  \centering
  \subfloat{
      \includegraphics[width=3.3in]{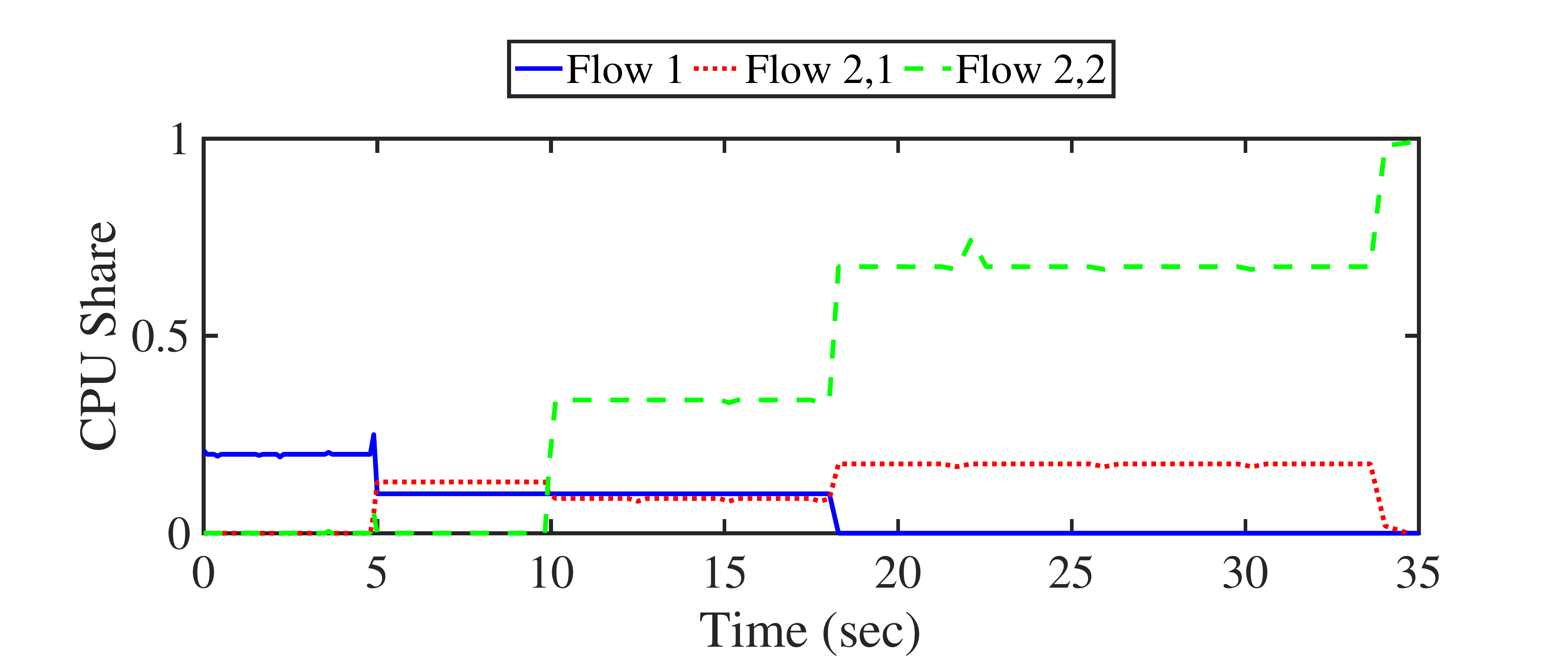}
      \label{fig_5:subfig:a}}
      \hfil
  \subfloat{
      \includegraphics[width=3.3in]{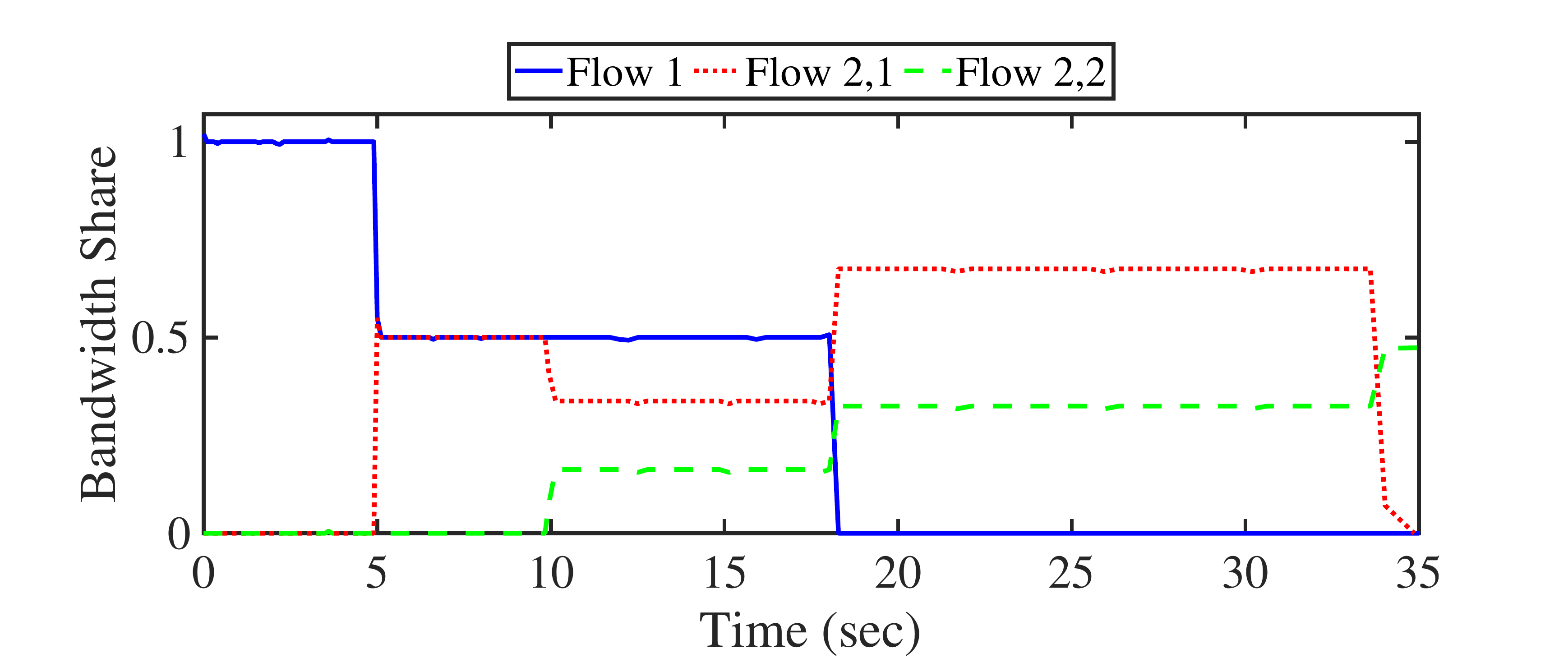}
      \label{fig_5:subfig:b}}
      \hfil
  \subfloat{
      \includegraphics[width=3.3in]{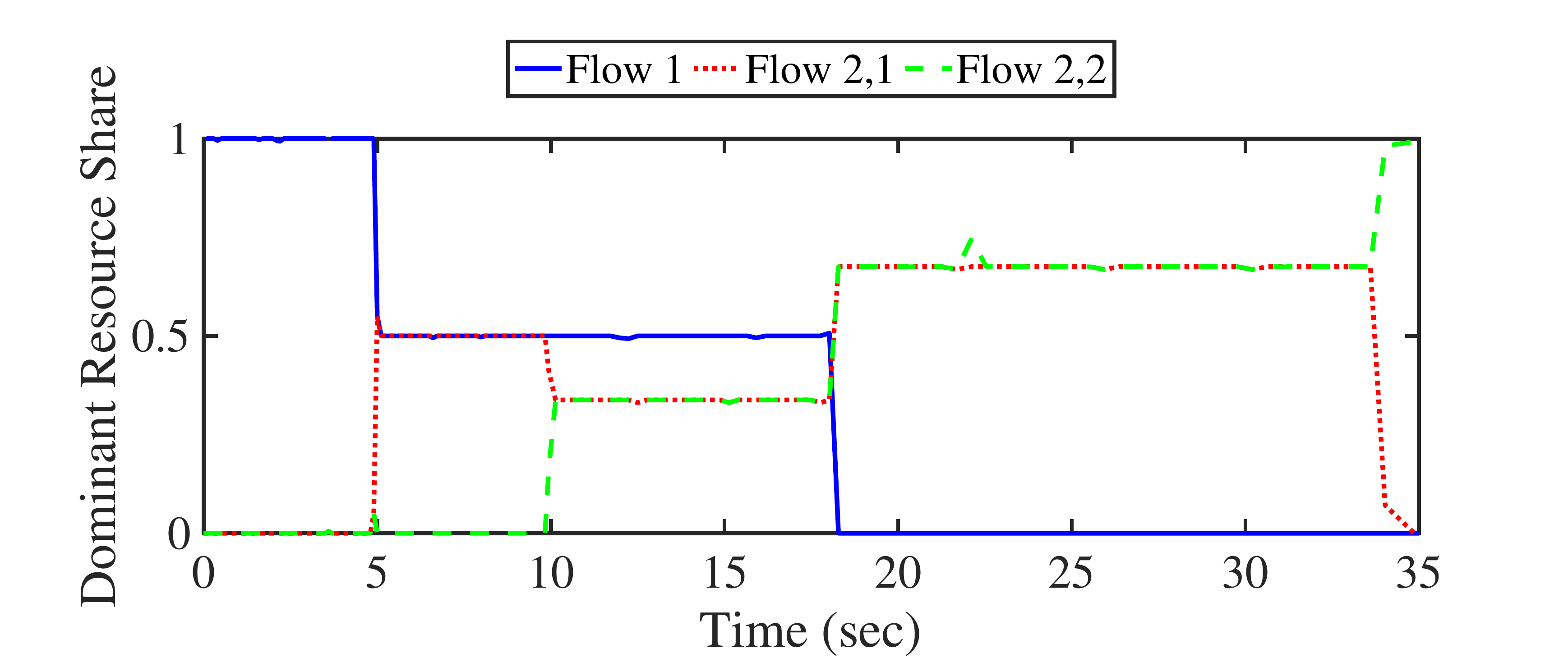}
      \label{fig_5:subfig:c}}
  \caption{Shares of three competing hierarchical flows arriving at different times. Flow 1, Flow 2,1 and Flow 2,2 respectively undergo basic forwarding, statistical monitoring and IPSec.}
  \label{fig_5}
\end{figure}
\begin{figure}[!t]
\centering
  \includegraphics[width=3.5in]{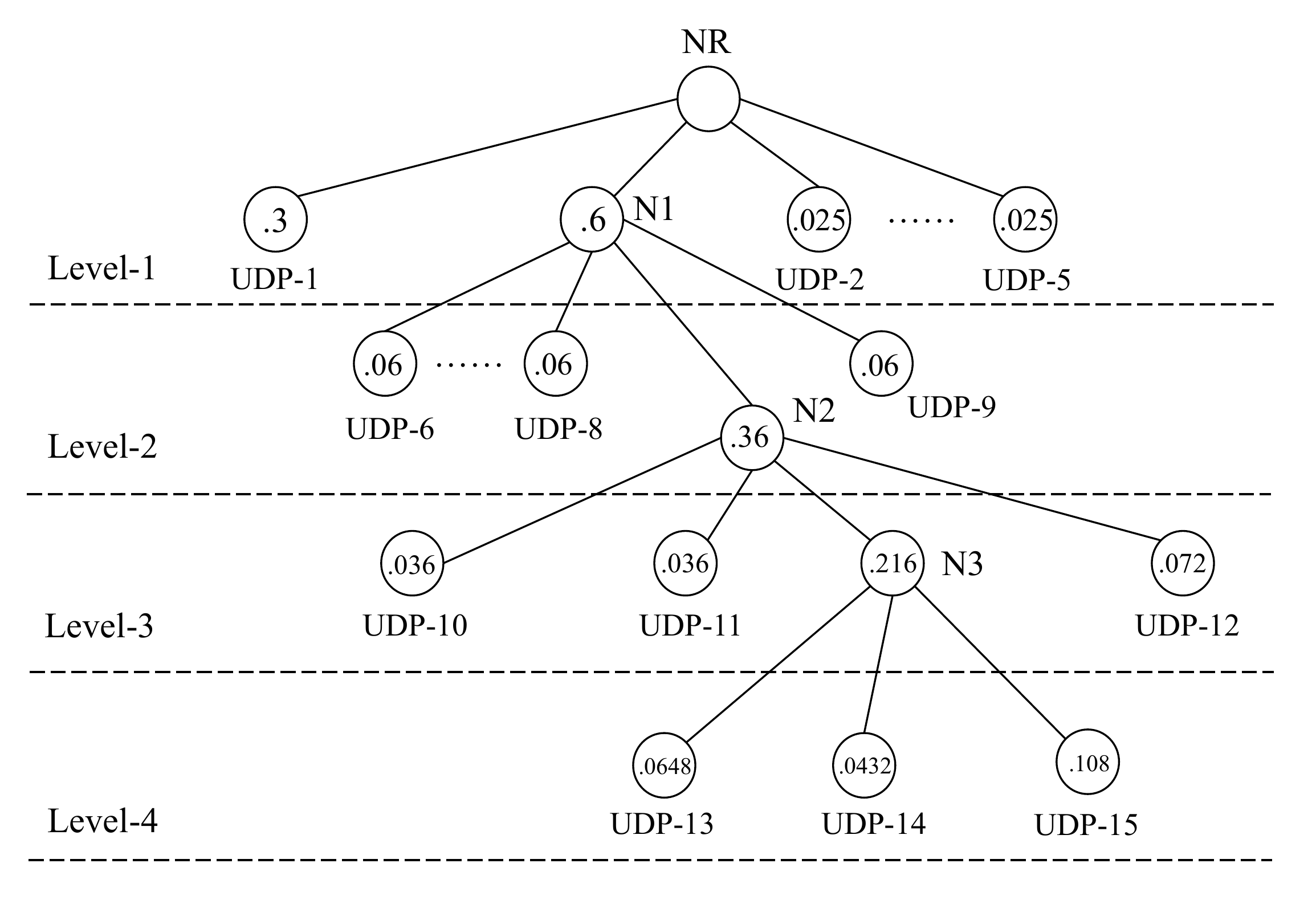}
  \caption{Hierarchy used in Section \ref{sec:5.2}}
  \label{fig_6}
\end{figure}

Fig. \ref{fig_5} shows the dynamic resource share allocated to the three physical flows over time under the two H-DRFQ algorithms proposed in our work.
Although the scheduling orders of packets in the two algorithms are not exactly the same, both algorithms provide the same results of resource shares allocated to each flow in a long time perspective.
Since $f_1$ is the only active flow during time interval $[0,5]$, it receives $20\%$ of the CPU share and all the bandwidth.
During $[5,10]$, $f_{1}$ and $f_{2,1}$ are active.
Although the weight of $f_{2,1}$ is 0.25, at this point it has no sibling nodes.
Accordingly, the resources that are supposed to be allocated to $f_{2,2}$ are allocated to $f_{2,1}$.
In other words, the weight of $f_{2,1}$ during $[5,10]$ can be regarded as 0.5.
As a result, the dominant shares of $f_1$ and $f_{2,1}$ in $[5,10]$ are the same.
Meanwhile, their bandwidth shares are also the same as they are both bandwidth bottleneck.
Later, when $f_{2,2}$ becomes active in $10$ s, all the three flows are backlogged during the time interval $[10,17]$.
Since $f_{2,1}$ and $f_{2,2}$ are a pair of sibling nodes, their dominant shares are supposed to be the same, as the figure indicates.
Since $f_{2,2}$ is CPU-bound, it grabs only $33.3\%$ of the bandwidth allocated to $f_2$ (\emph{i.e.}, $16.7\%$ of the total bandwidth), while $66.7\%$ of the CPU resources are allocated to $f_2$.
Meanwhile, since the arrival of $f_{2,2}$ only leads to the update of the resource allocation of $f_2$, the CPU share, bandwidth share, dominant share of $f_{1}$ remain unchanged until $f_1$ finishes in around $17$ s.
Similar DRFQ allocations are observed between each pair of sibling nodes in subsequent time intervals. Through the whole process, we see that H-DRFQ algorithms can quickly adapt to traffic dynamics, while providing predictable hierarchical share guarantees cross flows.

\section{Performance Evaluation}\label{sec:8}

We conduct simulation experiments using a more complicated hierarchy shown in Fig.~\ref{fig_6}, with aim to (1) examine the difference in scheduling delay between collapsed H-DRFQ and dove-tailing H-DRFQ and the sensitivities of delays with respect to the weights of flows;
(2) show the comparison of runtime between collapsed H-DRFQ and dove-tailing H-DRFQ.

\subsection{General Setup} \label{sec:8.1}

We implement 3 schedulers, collapsed H-DRFQ, dove-tailing H-DRFQ and DRFQ. By using the DRFQ scheduler we remove hierarchy shown in Fig.~\ref{fig_6} but keep the weights of the leaves, thus the DRFQ scheduler schedules the leaves according to their weights shown in the figure.
All simulation results are based on our event-driven packet simulator realized with around 3,000 lines of python codes. By default, packets follow Poisson arrivals.

We generate 15 UDP flows that sequentially start every 0.1s, as there are 15 leaves in the hierarchy shown in Fig.~\ref{fig_6}. A flow randomly chooses one of the three middlebox modules to pass through, and the middlebox module a flow passes through keeps the same across different algorithms. Moreover, the order of flows entering the leaves keeps the same across different algorithms. To congest the VNF resources, each flow sends $100,000$ UDP packets per second and the packet size is uniformly distributed between $[200, 1400]$ Bytes.

\subsection{Delay Comparison} \label{sec:8.2}
\begin{figure}[!t]
  \centering
  \includegraphics[width=2.0in]{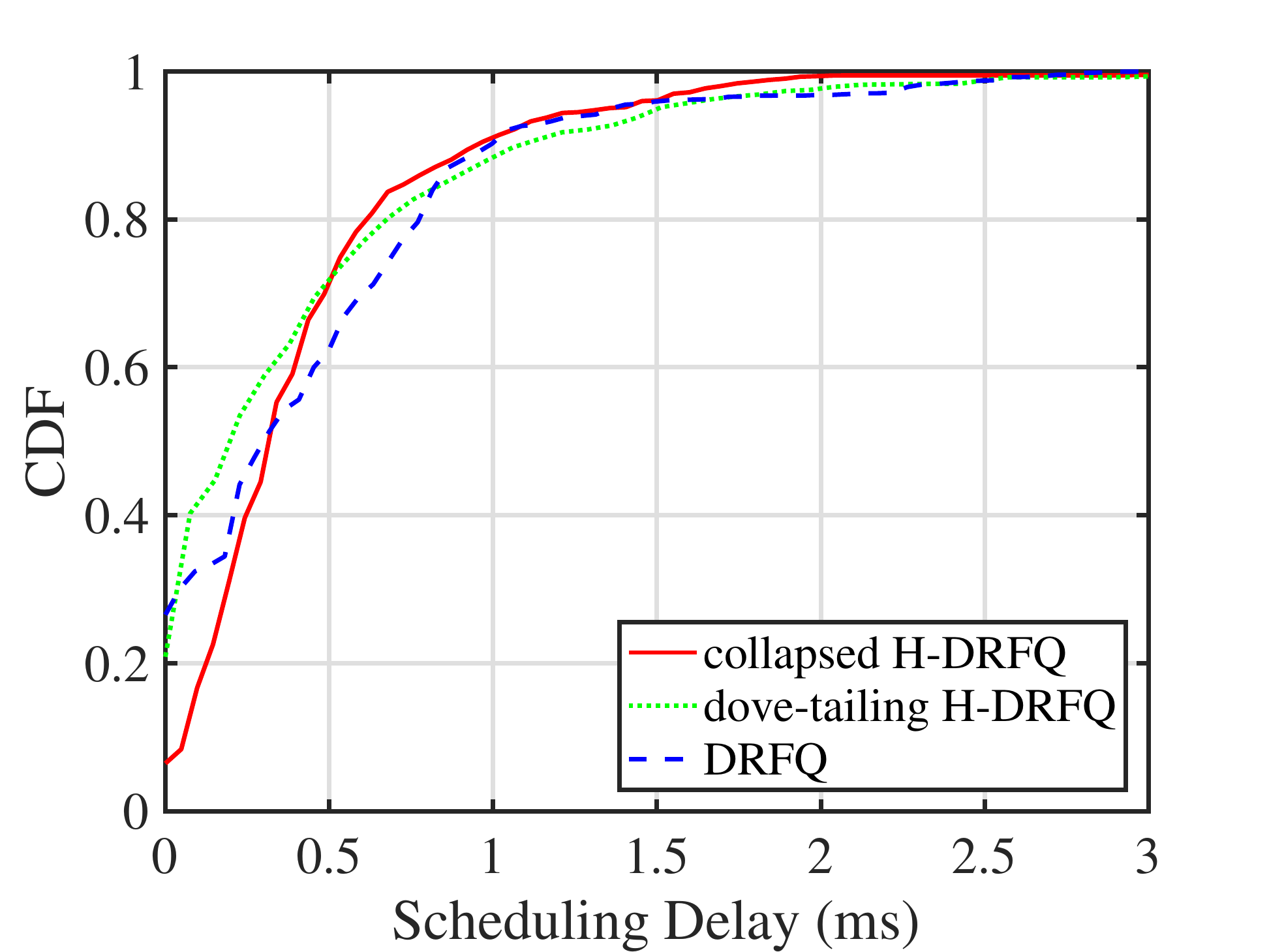}
  \caption{CDF of the scheduling delay}
  \label{fig_7}
\end{figure}

We first compare the packet scheduling delay under the three schedulers. Fig.~\ref{fig_7} depicts the CDF of the scheduling delay a packet experiences, from which we see that all packets can be scheduled in less than $6$ ms after their arrivals, among which nearly $80\%$ of the packets can be scheduled in $0.8$~ms.
Further investigation reveals, around $20\%$ of the packets under dove-tailing H-DRFQ experience short delays that approximate to zero, however, for collapsed H-DRFQ, this number becomes $8\%$. Nevertheless, nearly all packets under collapsed H-DRFQ start their services within $2$ ms, however, there are still $2\%$ of the packets under dove-tailing H-DRFQ waiting for service within the same time period. This result suggest that the short delays experienced by more packets under dove-tailing H-DRFQ are achieved at the expense that some of its packets have to suffer long delays. Unlike the theoretical result given by Corollary~\ref{cor:2} that dove-tailing H-DRFQ always outperforms collapsed H-DRFQ in terms of a smaller delay bound, actual delays tell us there is no permanent winner for the two H-DRFQ algorithms.
These experimental results are reasonable because the only scheduling difference between collapsed and dove-tailing H-DRFQ is the packet order, and the packet scheduling delay reduction of some nodes in a hierarchy must be paid back by the other nodes.

\begin{figure}[!t]
  \centering
  \subfloat[Per-level mean delay comparison]{
      \includegraphics[width=1.7in]{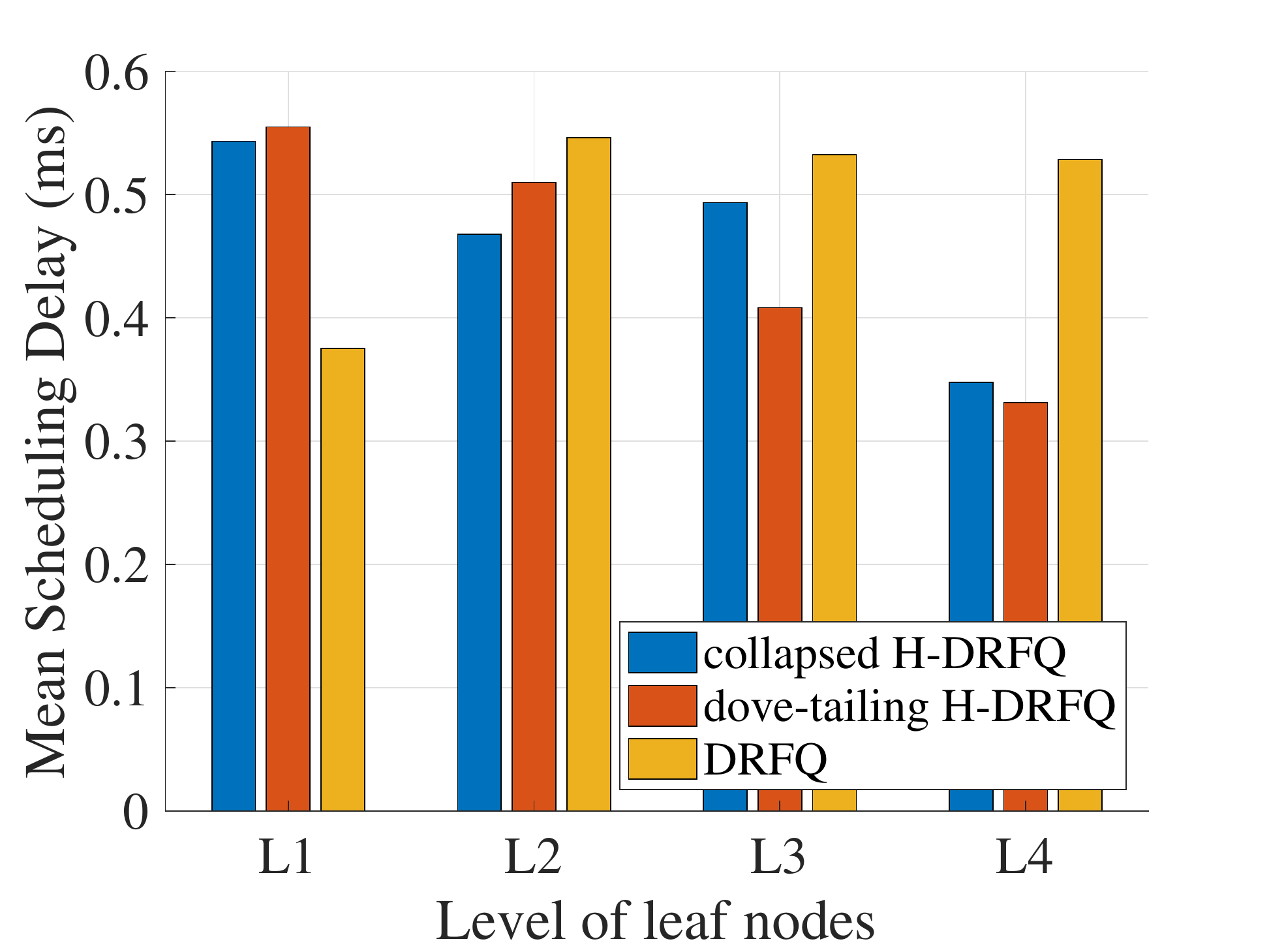}
      \label{fig_10:subfig:a}}
  \subfloat[Per-level maximum delay comparison]{
      \includegraphics[width=1.7in]{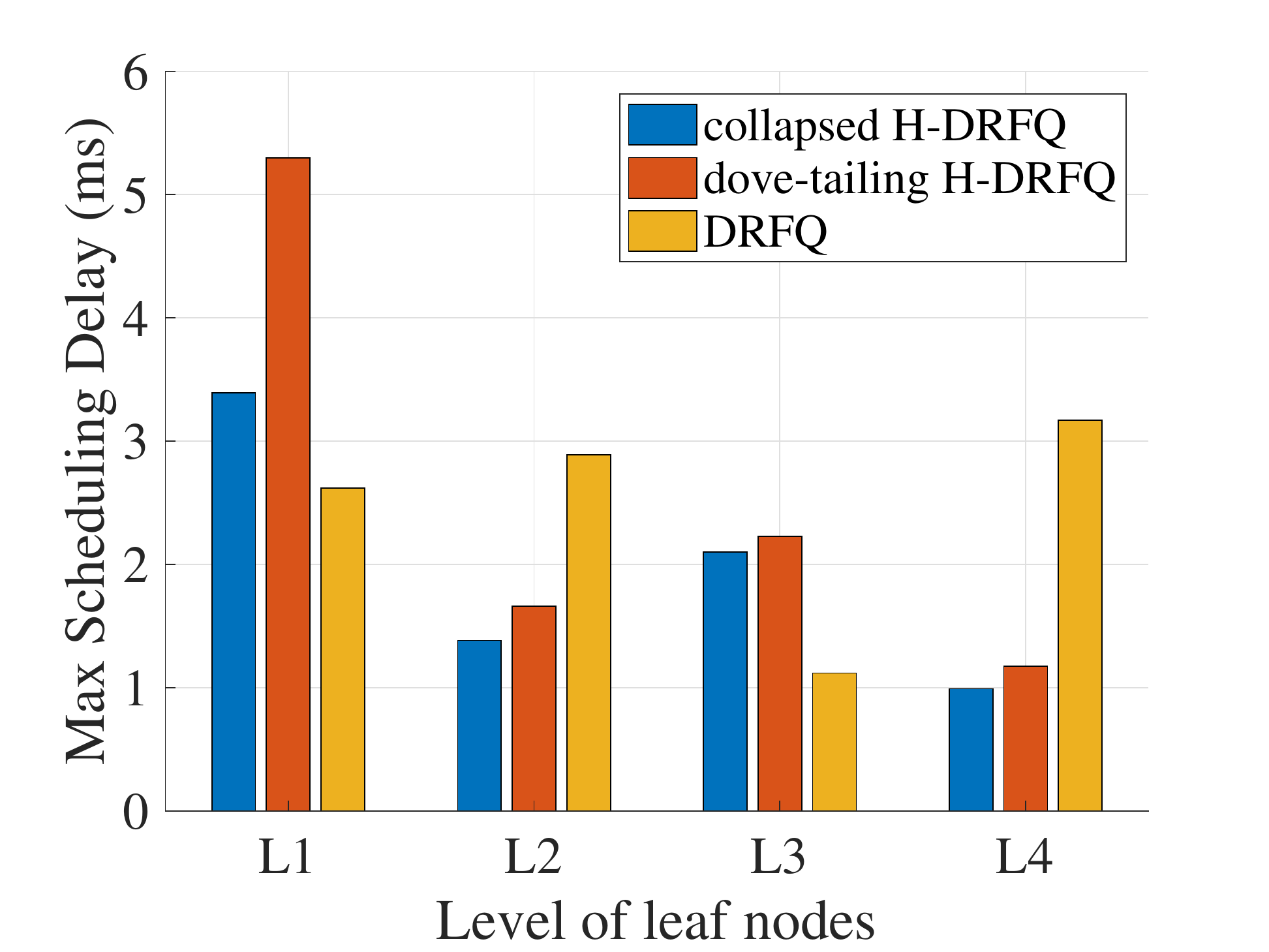}
      \label{fig_10:subfig:b}}
  \caption{Comparison of scheduling delay of collapsed and dove-tailing H-DRFQ under the hierarchy shown in Fig. \ref{fig_6}.}
  \label{fig_10}
\end{figure}

A detailed statistical result is given in Fig.~\ref{fig_10}(a) and~\ref{fig_10}(b). These two figures show the average and maximal delays a packet experiences in different levels, respectively.
We observe that the mean packet delays of nodes from $L_2$, $L_3$ and $L_4$ under DRFQ are longer than those under the H-DRFQ algorithms, however, the packet delays from $L_1$ show the opposite results.
This indicates that leaves benefit (experience shorter packet delays) from having longer distances to their root.

More importantly, Fig.~\ref{fig_10}(a) shows that in general the lower the level is (from L1 to L4), the shorter the delay a packet experiences.
Particularly, the packets under collapsed H-DRFQ experience shorter packet delays at higher levels (\emph{i.e.}, L1 and L2), while the packets under dove-tailing H-DRFQ experience shorter delays at lower levels (\emph{i.e.}, L3 and L4).
We attribute this phenomenon to the fact that in collapsed H-DRFQ, after the hierarchy is flattened, the weight of a node from levels other than L1 will always be larger than its original weight but smaller than the weight of its parent.
Consider the example hierarchy shown in Fig.~\ref{fig_3}, after the hierarchy is flattened, the weight of $f_{2,1}$ is updated to $0.33$, larger than its original weight $0.25$, but smaller than its parent $f_2$'s weight $0.5$. Meanwhile, the weight of $f_1$ remains the same.
According to the flow scheduling principle of memoryless DRFQ, this weight change leads to an increase in scheduling granularity of the nodes from the levels other than L1.
Therefore, compared with dove-tailing H-DRFQ, in collapsed H-DRFQ, the longer the distance of a node from the root, its next packet that needs to be scheduled is more likely to be delayed by the one from the other node that has shorter distance from the root.

\begin{figure*}[!t]
\centering
  \subfloat[Mean scheduling delay in L1]{
      \includegraphics[width=1.7in]{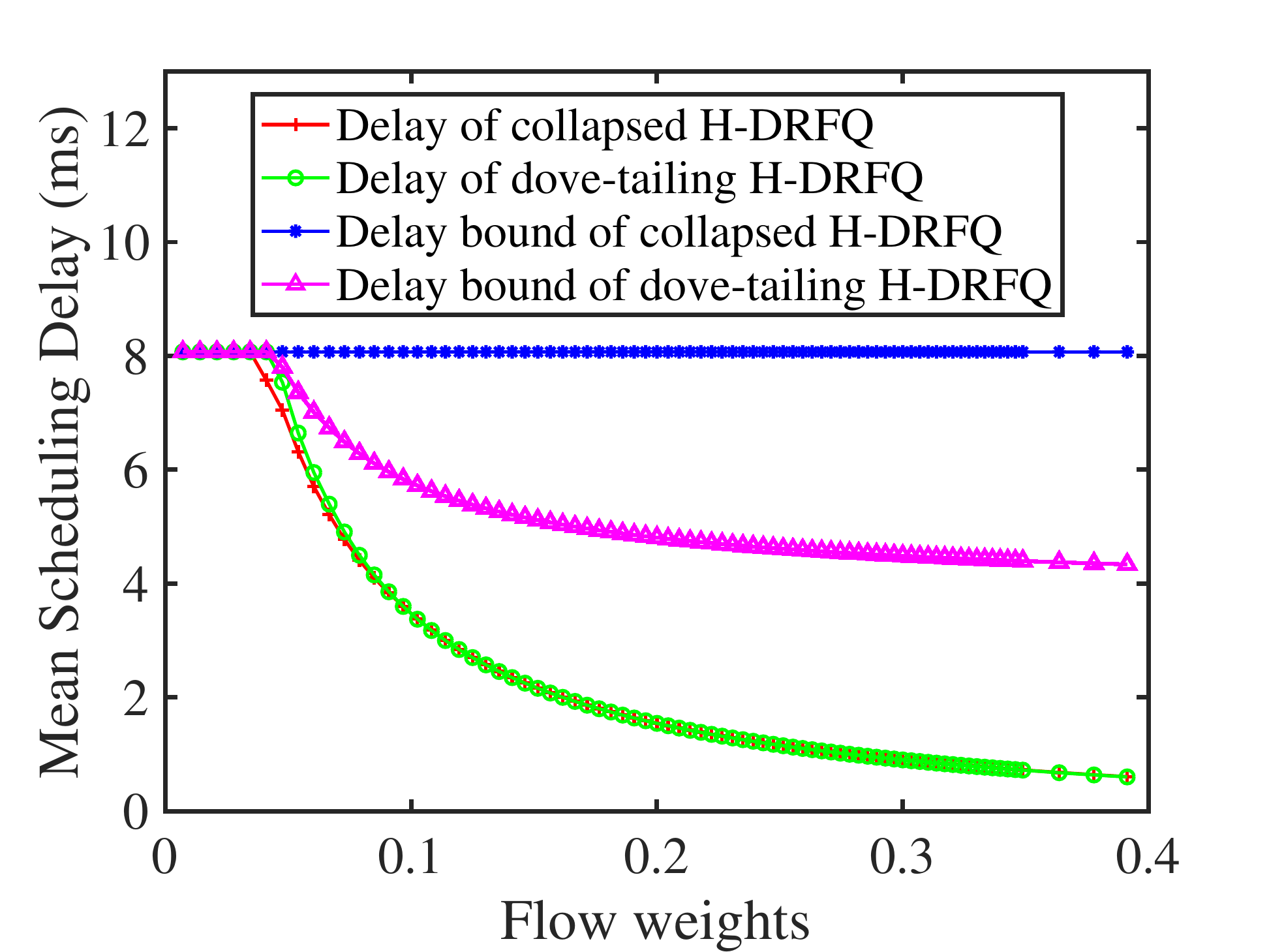}
      \label{fig_8:subfig:a}}
  \subfloat[Mean scheduling delay in L2]{
      \includegraphics[width=1.7in]{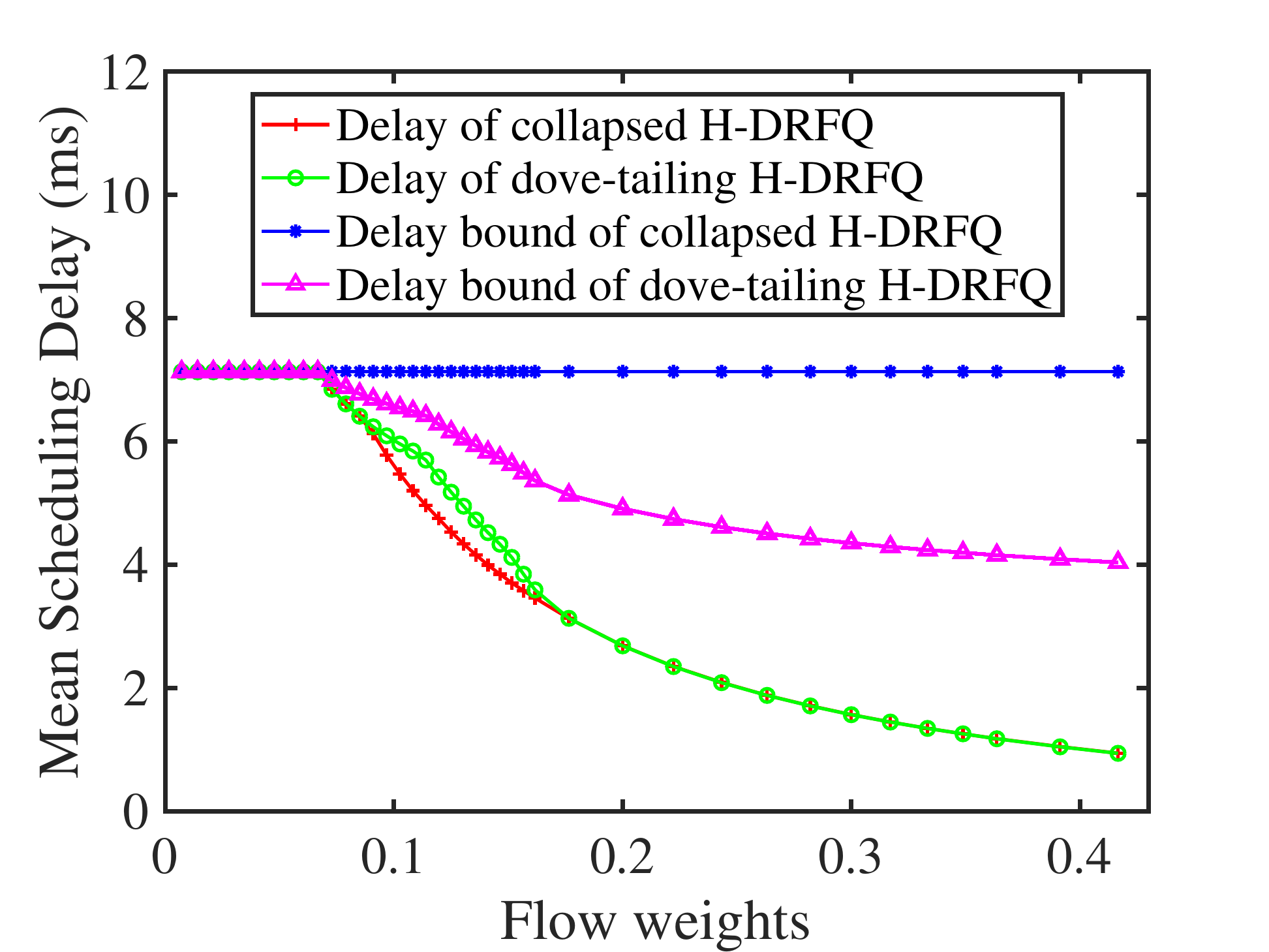}
      \label{fig_8:subfig:b}}
  \subfloat[Mean scheduling delay in L3]{
      \includegraphics[width=1.7in]{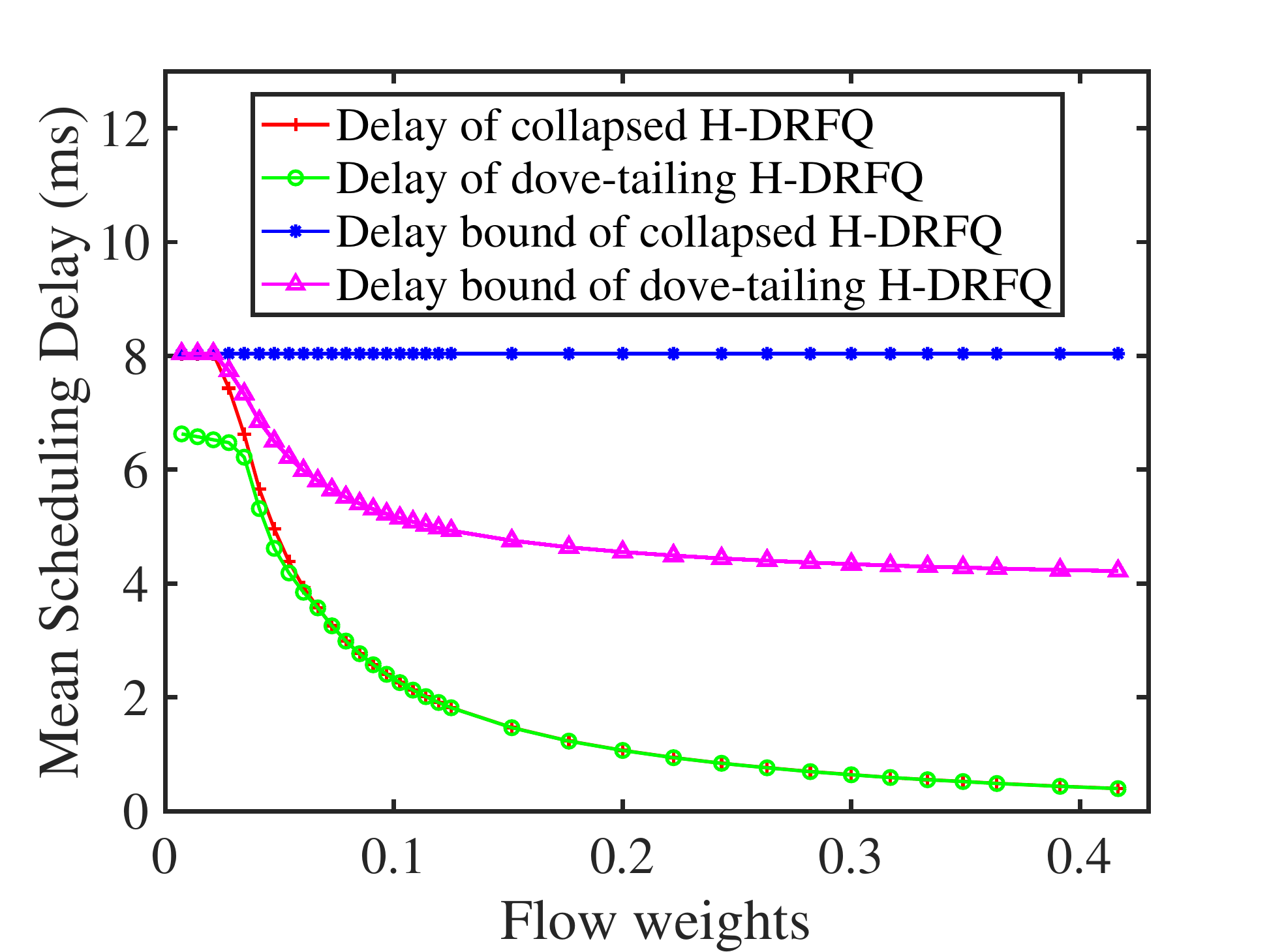}
      \label{fig_8:subfig:c}}
  \subfloat[Mean scheduling delay in L4]{
      \includegraphics[width=1.7in]{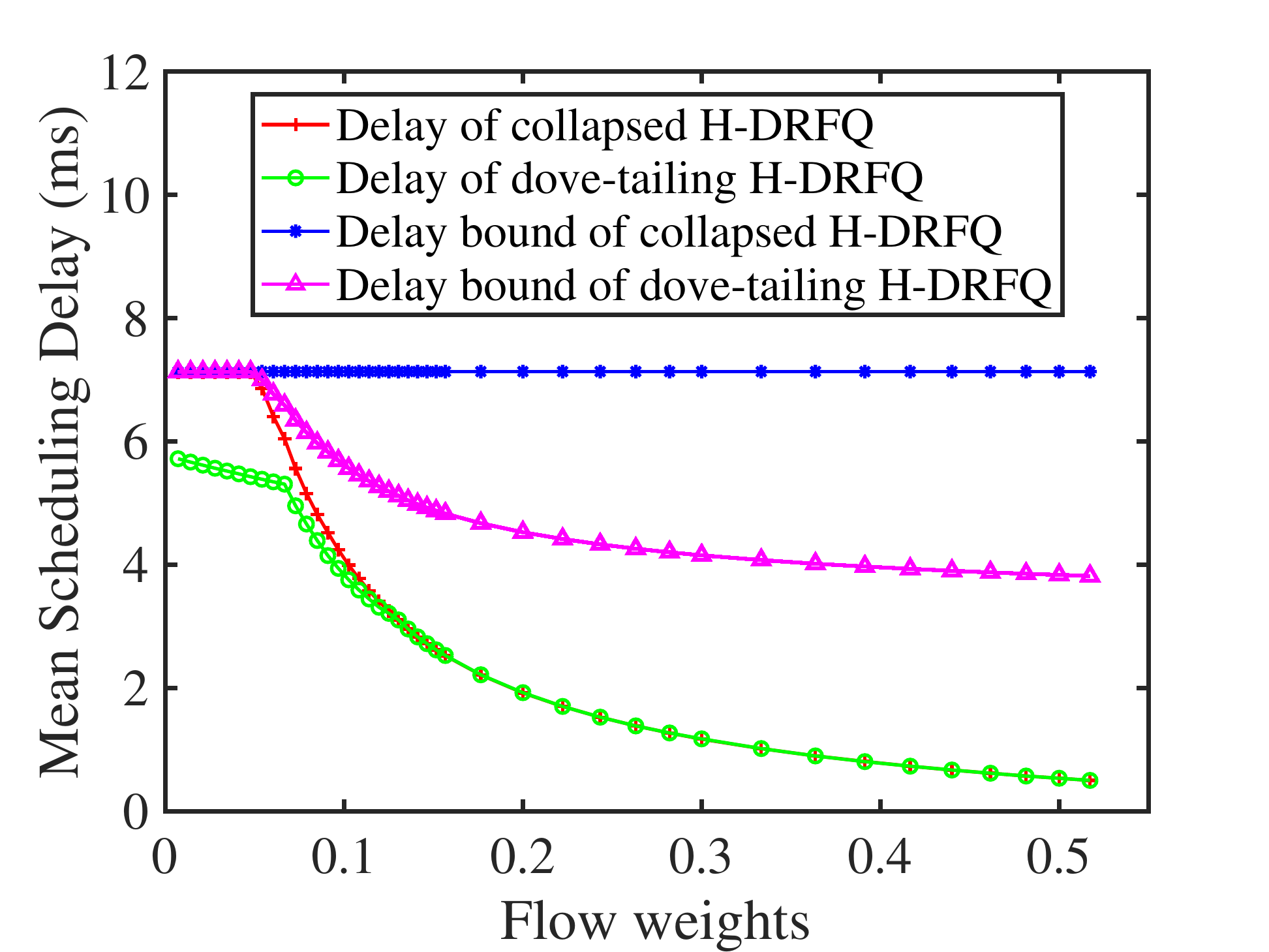}
      \label{fig_8:subfig:d}}
  \caption{Scheduling delay comparison between collapsed and dove-tailing H-DRFQ with respect to weights.}
  \label{fig_8}
\end{figure*}

\vspace{0.05in}
\noindent\textbf{Weight Sensitivity}. Next we change the weights of the nodes in the hierarchy. Fig.~\ref{fig_8} shows the mean scheduling delay a flow experiences with respective to its weight in each of the four levels in the hierarchy.
We observe that delay decreases as the weight increases.
Still, the packets scheduled by collapsed H-DRFQ experience shorter delays in L1 and L2, while the packets scheduled by dove-tailing H-DRFQ experience shorter delays in L3 and L4.
Moreover, with the increase of weights, the scheduling delays experienced by the same packets tend to be the same for the two H-DRFQ algorithms.
Interestingly, the delay bound of the collapsed H-DRFQ remains the same as the weight increases. However, the delay bound of the dove-tailing H-DRFQ decreases as the weight increases, showing the same tendency with the actual delays packets experience.
As the theoretical analysis in Section~\ref{sec:6.4} indicates, the upper bound of $D^c(p_i^k)$ is not tight. Each of the two H-DRFQ algorithms has its own advantage with respect to the level of flows in the hierarchy.

\subsection{Runtime Comparison}\label{sec:8.3}

Next we compare the total runtime of the 3 schedulers under different arrival patterns and packet sizes.
We measure the runtime of the schedulers using the hierarchy shown in Fig.~\ref{fig_6}, where runtime is used to reflect the time complexity.
We compare the runtime of the schedulers with respect to large, small and random packet sizes, under constant and exponential interarrival time respectively.

Table~\ref{tab:1} shows the total runtime of the three algorithms under different arrival patterns and packet sizes. From this table, we observe that while the runtime is highly consistent under different arrival patterns using the DRFQ scheduler, it is affected by the arrival patterns using the H-DRFQ scheduler. The runtime under the exponential interarrival traffic pattern is larger than that under the constant interarrival pattern. Moreover, the packet size also slightly affects the runtime. Therefore, although both the traffic arrival pattern and the packet size distribution are not included in the expressions of theoretical time complexity analysis, they indeed affect the practical runtime.

Another observation is that the runtime under dove-tailing H-DRFQ is about 3 to 5 times larger than the runtime under collapsed H-DRFQ, while the runtime under DRFQ is rather small. This is not in accordance with the time complexity analysis, as the time complexity under collapsed H-DRFQ, dove-tailing H-DRFQ and DRFQ is $\mathcal{O}(19)$, $\mathcal{O}(\log 5+\log 4 +\log 3 + \log 3)$ and $\mathcal{O}(\log 15)$ respectively.
Nevertheless, those experimental results shown in Table~\ref{tab:1} are still reasonable. This is because, in practice, only when a new flow arrives the collapsed H-DRFQ scheduler needs to flatten the hierarchy. Therefore, although the theoretical time complexity of collapsed H-DRFQ is $\mathcal{O}(19)$, which is the time complexity of flattening a hierarchy, its actual time complexity is close to $\mathcal{O}(\log 15)$, which is the time complexity of DRFQ scheduling after the hierarchy is flattened. Therefore, the actual runtime of collapsed H-DRFQ should be larger than that of DRFQ, but much smaller than that of dove-tailing H-DRFQ.

Overall, the runtime of the two H-DRFQ algorithms is much larger than that of DRFQ, but still within acceptable limits. Given that we use a 4-level hierarchy with 15 UDP flows that arrives dynamically, the hierarchy we considered in this paper is pretty complicated. In practice, as long as the hierarchy is not too complicated, the two H-DRFQ algorithms are always wise choices to process hierarchical scheduling.

\begin{table}[!t]
\centering
\caption{Total runtime of collapsed H-DRFQ, dove-tailing H-DRFQ and DRFQ under different arrival patterns and packet sizes (s).\label{tab:1}}
{\scriptsize
\begin{tabular}{|c|c|c|c|c|c|c|}
  \hline
  \text{ } & \multicolumn{3}{c|}{constant interarrival} & \multicolumn{3}{c|}{exponential interarrival} \\ \hline
  \text{ } & large & small & random & large & small & random \\ \hline
  \tabincell{c}{Collapsed\\H-DRFQ} & 2.22 & 2.73 & 3.06 & 2.55 & 3.19 & 3.28 \\ \hline
  \tabincell{c}{Dove-tailing\\H-DRFQ} & 7.47 & 8.60 & 9.48 & 13.18 & 11.70 & 14.12 \\ \hline
  DRFQ & 0.42 & 0.34 & 0.38 & 0.45 & 0.48 & 0.40 \\ \hline
\end{tabular}}
\end{table}

\section{Related Works}

Traditionally, for a single resource and flat scheduling, many classic fair queueing algorithms, such as WFQ~\cite{demers1989analysis}, GPS~\cite{parekh1993generalized}, DRR~\cite{shreedhar1995efficient}, and SFQ~\cite{goyal1996start}, were proposed to provide share guarantees.
In this section we present further details on most closely related works in the literature.
We categorize these works into two classes: multi-resource flat fair queueing and single-resource hierarchical fair queueing.

\textbf{\emph{Multi-resource Flat Fair Queueing}}: Multi-resource flat fair queueing has been vastly studied recently.
Notably, Ghodisi \emph{et al}. proposed Dominant Resource Fair Queueing (DRFQ) that achieves Dominant Resource Fairness (DRF)~\cite{ghodsi2011dominant} among ungrouped flows.
It ensures that the flows receive roughly the same packet processing time on their respective \emph{dominant resources, i.e.}, the resources they respectively require the most processing time on.
From DRFQ, a string of flow-up papers~\cite{wang2013multi2},~\cite{wang2013multi},~\cite{wang2014low},~\cite{li2015low},~\cite{chen2017cluster} modified, developed and extended DRFQ.
However, their focus has so far been primarily on \emph{flat} or \emph{non-hierarchical} flow scheduling. Our work bridges this gap and thus complements existing studies.

\textbf{\emph{Single-resource Hierarchical Fair Queueing}}:
The hierarchical structures of traffic flows are common in any type of networks~\cite{chowdhury2016hug},~\cite{tufail2016service}.
To provide hierarchical share guarantees, several hierarchical scheduling algorithms were proposed in \cite{goyal1996hierarchical}, \cite{bennett1997hierarchical}, \cite{stoica2000hierarchical}, \cite{chandra2008hierarchical}. These works extend the fair queueing problem from one-level flat tree to a multi-level hierarchy.
However, none of these works consider multiple resources.

Recently, Bhattacharya \emph{et al}. proposed a fair sharing algorithm, named Hierarchical Dominant Resource Fairness (HDRF)~\cite{bhattacharya2013hierarchical}, that supports hierarchical fair resource scheduling. However, like DRF, HDRF is also designed for sharing multiple types of resources in \emph{space} domain.
Given a cluster with much larger number of servers than users, the resource sharing in space domain decides how many resources a user should get on each server. In contrast, a middlebox scheduler requires resource sharing in \emph{time}. Given a small amount of resources (\emph{e.g.}, CPUs and GPUs) that each can only process one packet at a time, the middlebox scheduler has to interleave packets to receive appropriate resource shares over time.
Therefore, HDRF is not applicable to a middlebox scheduler where multiple types of resources are shared in time domain.
Our work, on the other hand, designs an algorithm that supports hierarchical fair queueing across multiple resources in time domain.

\section{Conclusions}

In this paper, we have studied the problem of hierarchical multi-resource fair queueing in a middlebox, such that the QoS of grouped flows passing through it can be guaranteed. We have proposed two new hierarchical multi-resource fair queueing algorithms, collapsed H-DRFQ and dove-tailing H-DRFQ. Collapsed H-DRFQ converts the hierarchy into a flat tree and then uses the flat tree as a DRFQ scheduler, while dove-tailing H-DRFQ applies dove-tailing DRFQ between each pair of sibling nodes.
Both algorithms can provide hierarchical share guarantees.
Theoretical analysis indicates that the delay bound of dove-tailing H-DRFQ is smaller than that of collapsed H-DRFQ, while experimental results shows that the actual packet delays vary with the position of the flow that possess the packet in the hierarchy. Both H-DRFQ algorithms are general methods in providing hierarchical share guarantees to grouped flows across multiple types of resources. With the surge of user data and the large-scale deployments of middlboxes, the two H-DRFQ algorithms could be used for effectively providing fine-grained QoS service to the data flows with the rapid growth of extremely diversified service demands.

\ifCLASSOPTIONcaptionsoff
  \newpage
\fi
\bibliographystyle{IEEEtran}
\bibliography{IEEEabrv,IEEEexample}

\begin{IEEEbiography}[{\includegraphics[width=1in,height=1.25in,clip,keepaspectratio]{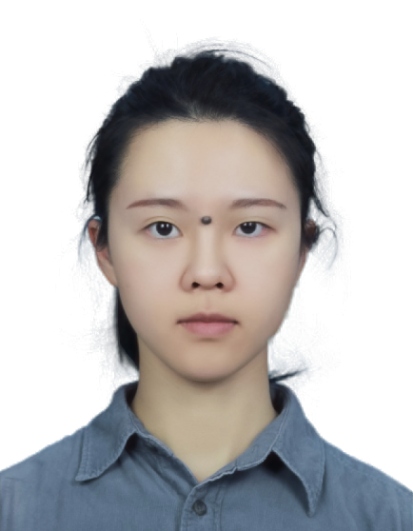}}]{Chaoqun You}
is a postdoctoral research fellow in Singapore University of Technology and Design (SUTD). She received the B.S. degree in communication engineering and the Ph.D. degree in communication and information system from University of Electronic Science and Technology of China (UESTC) in 2013 and 2020, respectively. Her research interests include mobile edge computing, network virtualzations, federated learning and meta-learning.
\end{IEEEbiography}
\begin{IEEEbiography}[{\includegraphics[width=1.2in,height=1.35in,clip,keepaspectratio]{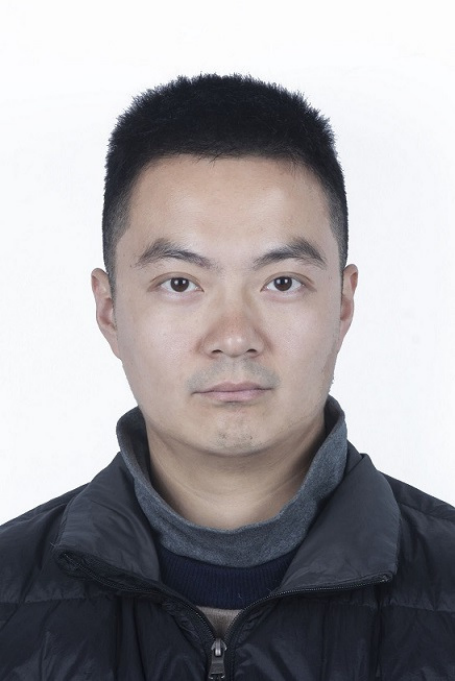}}]{Yangming Zhao}
is a research professor with the School of Computer Science and Technology, University of Science and Technology of China. Prior to that, he was a research scientist with University at Buffalo. He received the B.Eng. degree in communication engineering and the Ph.D. degree in communication and information system from University of Electronic Science and Technology of China in 2008 and 2015, respectively. His research interests include optimization related topics in data communication networks, such as data center networks, edge computing, transportation systems, and quantum networks.
\end{IEEEbiography}
\begin{IEEEbiography}[{\includegraphics[width=1in,height=1.25in,clip,keepaspectratio]{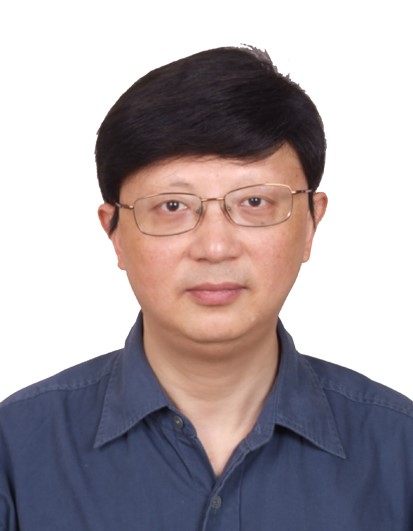}}]{Gang Feng}
(M’01, SM’06) received his BEng and MEng degrees in Electronic Engineering from the University of Electronic Science and Technology of China (UESTC), in 1986 and 1989, respectively, and the Ph.D. degrees in Information Engineering from The Chinese University of Hong Kong in 1998. He was with the School of Electric and Electronic Engineering, Nanyang Technological University in December 2000 as an assistant professor and an associate professor in October 2005. At present he is a professor with the National Laboratory of Communications, UESTC. Dr. Feng has extensive research experience and has published widely in computer and wireless networking research. A number of his papers have been highly cited. He has received the IEEE ComSoc TAOS Best Paper Award and ICC best paper award in 2019. His research interests include next generation mobile networks, mobile cloud computing, AI-enabled wireless networking, etc. Dr. Feng is a senor member of IEEE.
\end{IEEEbiography}

\begin{IEEEbiography}[{\includegraphics[width=1in,height=1.25in,keepaspectratio]{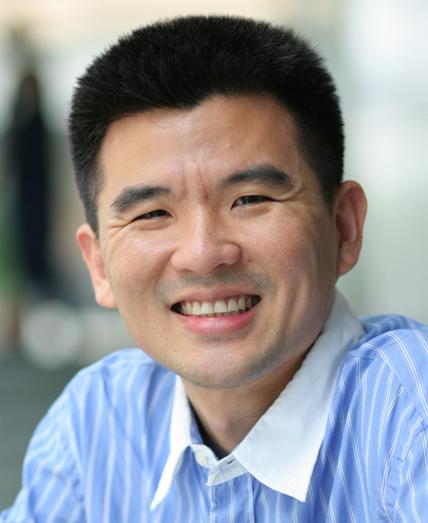}}]
{Tony Q.S. Quek}(S'98-M'08-SM'12-F'18) received the B.E.\ and M.E.\ degrees in electrical and electronics engineering from the Tokyo Institute of Technology in 1998 and 2000, respectively, and the Ph.D.\ degree in electrical engineering and computer science from the Massachusetts Institute of Technology in 2008. Currently, he is the Cheng Tsang Man Chair Professor with Singapore University of Technology and Design (SUTD). He also serves as the Director of the Future Communications R\&D Programme, the Head of ISTD Pillar, and the Deputy Director of the SUTD-ZJU IDEA. His current research topics include wireless communications and networking, network intelligence, internet-of-things, URLLC, and 6G.

Dr.\ Quek has been actively involved in organizing and chairing sessions, and has served as a member of the Technical Program Committee as well as symposium chairs in a number of international conferences. He is currently serving as an Area Editor for the {\scshape IEEE Transactions on Wireless Communications} and an elected member of the IEEE Signal Processing Society SPCOM Technical Committee. He was an Executive Editorial Committee Member for the {\scshape IEEE Transactions on Wireless Communications}, an Editor for the {\scshape IEEE Transactions on Communications}, and an Editor for the {\scshape IEEE Wireless Communications Letters}.

Dr.\ Quek was honored with the 2008 Philip Yeo Prize for Outstanding Achievement in Research, the 2012 IEEE William R. Bennett Prize, the 2015 SUTD Outstanding Education Awards -- Excellence in Research, the 2016 IEEE Signal Processing Society Young Author Best Paper Award, the 2017 CTTC Early Achievement Award, the 2017 IEEE ComSoc AP Outstanding Paper Award, the 2020 IEEE Communications Society Young Author Best Paper Award, the 2020 IEEE Stephen O. Rice Prize, the 2020 Nokia Visiting Professor, and the 2016-2020 Clarivate Analytics Highly Cited Researcher. He is a Fellow of IEEE.
\end{IEEEbiography}

\begin{IEEEbiography}[{\includegraphics[width=1in,height=1.25in,clip,keepaspectratio]{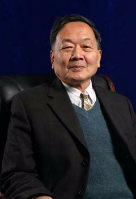}}]{Lemin Li}
received the BS degree in electrical engineering from the Jiaotong University, Shanghai, in 1952. Then, he was with the Department
of Electrical Communications, Jiaotong University until 1956. Since 1956, he has been with the Chengdu Institute of Radio Engineering (now the UESTC). During August 1980 to August 1982, he was a visiting scholar in the Department of Electrical Engineering and Computer Science, University of California at San Diego. His currently research interests are in the area of communication networks.
\end{IEEEbiography}

\end{document}